\documentstyle[twoside]{article}
\catcode`\@=11
\long\def\@makefntext#1{
\protect\noindent \hbox to 3.2pt {\hskip-.9pt
$^{{\eightrm\@thefnmark}}$\hfil}#1\hfill}		

\def\@makefnmark{\hbox to 0pt{$^{\@thefnmark}$\hss}}	
\def\ps@myheadings{\let\@mkboth\@gobbletwo
\def\@oddhead{\hbox{}
\rightmark\hfil\eightrm\thepage}
\def\@oddfoot{}\def\@evenhead{\eightrm\thepage\hfil
\leftmark\hbox{}}\def\@evenfoot{}
\def\sectionmark##1{}\def\subsectionmark##1{}}
\oddsidemargin=\evensidemargin
\newcounter{sectionc}\newcounter{subsectionc}\newcounter{subsubsectionc}
\renewcommand{\section}[1] {\vspace{12pt}\addtocounter{sectionc}{1}
\setcounter{subsectionc}{0}\setcounter{subsubsectionc}{0}\noindent
	{\tenbf\thesectionc. #1}\par\vspace{5pt}}
\renewcommand{\subsection}[1] {\vspace{12pt}\addtocounter{subsectionc}{1}
	\setcounter{subsubsectionc}{0}\noindent
	{\bf\thesectionc.\thesubsectionc. {\kern1pt \bfit #1}}\par\vspace{5pt}}
\renewcommand{\subsubsection}[1] {\vspace{12pt}\addtocounter{subsubsectionc}{1}
	\noindent{\tenrm\thesectionc.\thesubsectionc.\thesubsubsectionc.
	{\kern1pt \tenit #1}}\par\vspace{5pt}}
\newcommand{\nonumsection}[1] {\vspace{12pt}\noindent{\tenbf #1}
	\par\vspace{5pt}}
\newcounter{appendixc}\newcounter{subappendixc}[appendixc]
\newcounter{subsubappendixc}[subappendixc]
\renewcommand{\thesubappendixc}{\Alph{appendixc}.\arabic{subappendixc}}
\renewcommand{\thesubsubappendixc}
	{\Alph{appendixc}.\arabic{subappendixc}.\arabic{subsubappendixc}}
\renewcommand{\appendix}[1] {\vspace{12pt}
        \refstepcounter{appendixc}
        \setcounter{figure}{0}
        \setcounter{table}{0}
        \setcounter{lemma}{0}
        \setcounter{theorem}{0}
        \setcounter{corollary}{0}
        \setcounter{definition}{0}
        \setcounter{equation}{0}
        \renewcommand{\thefigure}{\Alph{appendixc}.\arabic{figure}}
        \renewcommand{\thetable}{\Alph{appendixc}.\arabic{table}}
        \renewcommand{\theappendixc}{\Alph{appendixc}}
        \renewcommand{\thelemma}{\Alph{appendixc}.\arabic{lemma}}
        \renewcommand{\thetheorem}{\Alph{appendixc}.\arabic{theorem}}
        \renewcommand{\thedefinition}{\Alph{appendixc}.\arabic{definition}}
        \renewcommand{\thecorollary}{\Alph{appendixc}.\arabic{corollary}}
        \renewcommand{\theequation}{\Alph{appendixc}.\arabic{equation}}
        \noindent{\tenbf Appendix \theappendixc #1}\par\vspace{5pt}}
\newcommand{\subappendix}[1] {\vspace{12pt}
        \refstepcounter{subappendixc}
        \noindent{\bf Appendix \thesubappendixc. {\kern1pt \bfit #1}}
	\par\vspace{5pt}}
\newcommand{\subsubappendix}[1] {\vspace{12pt}
        \refstepcounter{subsubappendixc}
        \noindent{\rm Appendix \thesubsubappendixc. {\kern1pt \tenit #1}}
	\par\vspace{5pt}}
\newcommand{\textlineskip}{\baselineskip=13pt}
\newcommand{\smalllineskip}{\baselineskip=10pt}
\def\eightcirc{
\begin{picture}(0,0)
\put(4.4,1.8){\circle{6.5}}
\end{picture}}
\def\eightcopyright{\eightcirc\kern2.7pt\hbox{\eightrm c}}
\newcommand{\copyrightheading}[1]
	{\vspace*{-2.5cm}\smalllineskip{\flushleft
	{\footnotesize International Journal of Modern Physics B, #1}\\
	{\footnotesize $\eightcopyright$\, World Scientific Publishing
	 Company}\\
	 }}
\newcommand{\pub}[1]{{\begin{center}\footnotesize\smalllineskip
	Received #1\\
	\end{center}
	}}

\def\abstracts#1#2#3{{
	\centering{\begin{minipage}{4.5in}\baselineskip=10pt\footnotesize
	\parindent=0pt #1\par
	\parindent=15pt #2\par
	\parindent=15pt #3
	\end{minipage}}\par}}

\renewenvironment{thebibliography}[1]			
	{\frenchspacing
	 \ninerm\baselineskip=11pt
	 \begin{list}{\arabic{enumi}.}
	{\usecounter{enumi}\setlength{\parsep}{0pt}
	 \setlength{\leftmargin 17pt}{\rightmargin 0pt}   
	 \setlength{\itemsep}{0pt} \settowidth
	{\labelwidth}{#1.}\sloppy}}{\end{list}}
\newcounter{itemlistc}
\newcounter{romanlistc}
\newcounter{alphlistc}
\newcounter{arabiclistc}
\newenvironment{itemlist}
    	{\setcounter{itemlistc}{0}
	 \begin{list}{$\bullet$}
	{\usecounter{itemlistc}
	 \setlength{\parsep}{0pt}
	 \setlength{\itemsep}{0pt}}}{\end{list}}
\newenvironment{romanlist}
	{\setcounter{romanlistc}{0}
	 \begin{list}{$($\roman{romanlistc}$)$}
	{\usecounter{romanlistc}
	 \setlength{\parsep}{0pt}
	 \setlength{\itemsep}{0pt}}}{\end{list}}

\newcommand{\fcaption}[1]{
        \refstepcounter{figure}
        \setbox\@tempboxa = \hbox{\footnotesize Fig.~\thefigure. #1}
        \ifdim \wd\@tempboxa > 5in
           {\begin{center}
        \parbox{5in}{\footnotesize\smalllineskip Fig.~\thefigure. #1}
            \end{center}}
        \else
             {\begin{center}
             {\footnotesize Fig.~\thefigure. #1}
              \end{center}}
        \fi}
\newcommand{\tcaption}[1]{
        \refstepcounter{table}
        \setbox\@tempboxa = \hbox{\footnotesize Table~\thetable. #1}
        \ifdim \wd\@tempboxa > 5in
           {\begin{center}
        \parbox{5in}{\footnotesize\smalllineskip Table~\thetable. #1}
            \end{center}}
        \else
             {\begin{center}
             {\footnotesize Table~\thetable. #1}
              \end{center}}
        \fi}
\def\@citex[#1]#2{\if@filesw\immediate\write\@auxout
	{\string\citation{#2}}\fi
\def\@citea{}\@cite{\@for\@citeb:=#2\do
	{\@citea\def\@citea{,}\@ifundefined
	{b@\@citeb}{{\bf ?}\@warning
	{Citation `\@citeb' on page \thepage \space undefined}}
	{\csname b@\@citeb\endcsname}}}{#1}}
\newif\if@cghi
\def\cite{\@cghitrue\@ifnextchar [{\@tempswatrue
	\@citex}{\@tempswafalse\@citex[]}}
\def\citelow{\@cghifalse\@ifnextchar [{\@tempswatrue
	\@citex}{\@tempswafalse\@citex[]}}
\def\@cite#1#2{{$\null^{#1}$\if@tempswa\typeout
	{IJCGA warning: optional citation argument
	ignored: `#2'} \fi}}
\newcommand{\citeup}{\cite}
\def\pmb#1{\setbox0=\hbox{#1}
	\kern-.025em\copy0\kern-\wd0
	\kern.05em\copy0\kern-\wd0
	\kern-.025em\raise.0433em\box0}

\def\fnm#1{$^{\mbox{\scriptsize #1}}$}
\def\fnt#1#2{\footnotetext{\kern-.3em
	{$^{\mbox{\scriptsize #1}}$}{#2}}}
\def\fpage#1{\begingroup
\voffset=.3in
\thispagestyle{empty}\begin{table}[b]\centerline{\footnotesize #1}
	\end{table}\endgroup}
\def\runninghead#1#2{\pagestyle{myheadings}
\markboth{{\protect\footnotesize\it{\quad #1}}\hfill}
{\hfill{\protect\footnotesize\it{#2\quad}}}}
\font\tenrm=cmr10
\font\tenit=cmti10
\font\tenbf=cmbx10
\font\bfit=cmbxti10 at 10pt
\font\ninerm=cmr9
\font\nineit=cmti9
\font\ninebf=cmbx9
\font\eightrm=cmr8

\def\qed{\hbox{${\vcenter{\vbox{			
   \hrule height 0.4pt\hbox{\vrule width 0.4pt height 6pt
   \kern5pt\vrule width 0.4pt}\hrule height 0.4pt}}}$}}
\def\bsc{{\sc a\kern-6.4pt\sc a\kern-6.4pt\sc a}}	
\def\bflatex{\bf L\kern-.30em\raise.3ex\hbox{\bsc}\kern-.14em
T\kern-.1667em\lower.7ex\hbox{E}\kern-.125em X}
\begin{document}
\runninghead{T. Portengen et al.}{Optics with Quantum Hall Skyrmions}
\normalsize\textlineskip
\thispagestyle{empty}
\setcounter{page}{1}
\copyrightheading{} 
\vspace*{0.88truein}
\fpage{1}
\centerline{\bf OPTICS WITH QUANTUM HALL SKYRMIONS}
\vspace*{0.37truein}
\centerline{\footnotesize T. PORTENGEN, J. R. CHAPMAN, V. NIKOS
NICOPOULOS, and N. F. JOHNSON}
\vspace*{0.015truein}
\centerline{\footnotesize\it
Department of Physics, University of Oxford, Parks Road}
\baselineskip=10pt
\centerline{\footnotesize\it Oxford, OX1 3PU, United Kingdom}
\vspace*{0.225truein}
\pub{\today}
\vspace*{0.21truein}
\abstracts{
     A novel type of charged excitation, known as a Skyrmion, has
   recently been discovered in quantum Hall systems with filling
    factor near $\nu = 1$. A Skyrmion---which can be thought of as
    a topological twist in the spin density of the electron
    gas---has the same charge as an electron, but a much larger spin.
    In this review we present a detailed theoretical
    investigation of the optical properties of Skyrmions. Our results
     provide means for the optical detection of Skyrmions using
    photoluminescence (PL) spectroscopy. We first consider the optical
      properties of Skyrmions in disordered systems. A calculation of
     the luminescence energy reveals a special optical signature which
    allows us to distinguish between Skyrmions and ordinary electrons.
    Two experiments to measure the optical signature are proposed.
     We then turn to the optical properties of Skyrmions in pure
    systems. We show that, just like an ordinary electron, a Skyrmion
    may bind with a hole to form a Skyrmionic exciton. The Skyrmionic
    exciton can have a lower energy than the ordinary magnetoexciton.
   The optical signature of Skyrmions is found to be a robust feature
    of the PL spectrum in both disordered and pure systems.}{}{}
\vspace*{1pt}\textlineskip
\section{Introduction}
\vspace*{-0.5pt}
\noindent
   Imagine piercing a sphere with arrows pointing in a direction
   normal to the surface of the sphere. Then imagine stereographically
  projecting the surface of the sphere onto a flat plane, keeping
   the direction of the arrows fixed. The resulting distribution
   of arrows in the plane is a vector field with a peculiar twist
   known as a Skyrmion.

   Skyrmions were introduced by T. H. R. Skyrme\citeup{Skyrme}
   in 1958 as a way of representing the nucleon in terms of the
   underlying pion field. During the following decades most interest
   for Skyrmions came from nuclear physicists and particle physicists.
   That changed a few years ago when S. L. Sondhi, A. Karlhede,
   S. A. Kivelson, and E. H. Rezayi\citeup{Sondhi}---building on
   earlier work by D. H. Lee and C. L. Kane\citeup{Lee}---predicted
    the existence of Skyrmions in a
   two-dimensional (2D) electron system subjected to a perpendicular
    magnetic field of a particular strength. The strength
   of the magnetic field is such that the filling factor $\nu = 2 \pi
   \ell^{2} n_{s}$ (where $\ell = \sqrt{\hbar c/e B}$ is the magnetic
   length and $n_{s}$ is the electron density) is equal to an odd
    integer ($\nu = 1,3,5,\ldots$). Under these conditions the
    2D electron system exhibits the well-known (integer) quantum
     Hall effect. Skyrmions occuring in a 2D electron
    system subjected to a magnetic field have thus become known as
     ``quantum Hall'' Skyrmions.

   The first experimental evidence for quantum Hall Skyrmions was
     obtained by Barrett, Dabbagh, Pfeiffer, West, and
    Tycko\citeup{Barrett}. They used nuclear magnetic resonance
    (NMR) spectroscopy to measure the Knight shift of $^{71}$Ga
   nuclei located in a $n$-doped GaAs quantum well. The Knight
    shift is directly proportional to the spin polarization of
   the 2D electron gas surrounding the nuclei. Barrett {\em et al.}
    observed a precipitous drop of the Knight shift as the filling
  factor is varied away from $\nu = 1$, indicating that the spin
   of the charged excitations at $\nu = 1$ is much larger than
   $\frac{1}{2}$. Evidence for Skyrmions was also obtained by
   Schmeller, Eisenstein, Pfeiffer, and West\citeup{Schmeller} using
   tilted-field magnetotransport measurements, and by Aifer, Goldberg,
   and Broido\citeup{Aifer} using interband optical transmission
   spectroscopy. Maude {\em et al.}\citeup{Maude} have determined
    the spin activation gap at $\nu = 1$ using transport measurements
  under hydrostatic pressure. They found that a spin activation gap
   exists even at zero Zeeman energy, in agreement with the theoretical
   prediction\citeup{Sondhi} based on Skyrmions.

   The past three years have seen an explosion of theoretical work
   on quantum Hall Skyrmions. Below we briefly summarize some important
  developments. In their pioneering paper Sondhi {\em et al.} employed
  a field-theoretic approach in which it was difficult to include
   the effect of a nonzero Zeeman energy. A variational Hartree-Fock
   approach which allows for the inclusion of the Zeeman energy was
   proposed by Fertig, Brey, C\^{o}t\'{e}, and MacDonald\citeup{Fertig}.
   These authors also proposed that the ground state away from
    odd-integer filling factors is a crystal of Skyrmions\citeup{Brey}.
   In addition, they studied the dynamics of a single
   Skyrmion bound to a charged impurity\citeup{Cote}. A field-theoretic
    approach to the low-temperature properties of a multi-Skyrmion
   system was developed by Green, Kogan, and Tsvelik\citeup{Green}.
   The variational wave functions for a single Skyrmion proposed by
   Fertig {\em et al}.\ are eigenstates of $L_{z} \pm S_{z}$ (where
   $L_{z}$ and $S_{z}$ are the $z$ components of the orbital and spin
   angular momentum, respectively), but not of $L_{z}$ and $S_{z}$
  separately. Skyrmion states that {\em are} eigenstates of $L_{z}$
   and $S_{z}$ were found by MacDonald, Fertig, and Brey\cite{MacDonald}
    for a hard-core model Hamiltonian. An alternative variational
   approach which uses Landau-gauge wave functions (rather than the
   circular-gauge wave functions used by Fertig {\em et al.}\citeup{Fertig})
   has been proposed by Bychkov, Maniv, and Vagner\citeup{Bychkov}.
   Numerical studies of Skyrmion excitations at $\nu = 1$ have been
    carried out by Xie and He\citeup{Xie}. The possible existence of
   Skyrmions at fractional filling factors (in particular
   $\nu = \frac{1}{3}$) has been explored by Kamilla, Wu, and
    Jain\citeup{Kamilla}, and by Ahn and Chang\citeup{Ahn}.

   In this review we give a detailed account of the optical properties
    of quantum Hall Skyrmions. In particular we address the question
    of how the optical properties of Skyrmions differ from those of
    ordinary electrons. Our aim is to provide means for the optical
    detection of Skyrmions using PL spectroscopy. Brief accounts of
   parts of this work have been published
     elsewhere\citeup{Portengen,Chapman}.

    The paper is organized as follows. In Sec.~2 we review the
    variational description of Skyrmions proposed by Fertig, Brey,
   C\^{o}t\'{e}, and MacDonald\citeup{Fertig}. This approach
       is very useful for discussing the optical properties
     of Skyrmions. We also review three key experiments that have
     given evidence for the existence of Skyrmions
       at $\nu = 1$. In Sec.~3 we give an overview of
     the PL experiments that have been carried out on
     quantum Hall systems, and recall what information these
    experiments provide. We reproduce the PL spectrum that is
      experimentally observed near $\nu = 1$ and explain its
     basic features in terms of a simple model. Sections~4
       and 5---which make up the main part of the paper---are
     concerned with the optical properties of quantum Hall
     Skyrmions in disordered and pure systems, respectively.
      The main results of the paper are summarized in Sec.~6.

\section{Skyrmions in Quantum Hall Ferromagnets}
  \noindent Consider a 2D electron gas lying in the $x$-$y$ plane
    and subjected to a perpendicular magnetic field ${\bf B} =
    B \hat{z}$. The physical properties of the 2D electron gas depend
    sensitively on the filling factor $\nu$. The filling factor is
     the ratio between the number
   of electrons and the orbital degeneracy of a Landau level. Here
    we consider systems for which the filling factor is an odd
     integer: $\nu = 1,3,5,\ldots$. We focus in particular on
   the filling factor $\nu = 1$ for which only the lowest Landau
    level is occupied.
     In the symmetric gauge with vector potential ${\bf A} =
    \frac{1}{2} B(x,-y,0)$, the single-particle
     wave functions for the lowest Landau level have the
    form\citeup{Chakraborty} (in units where $\ell =
     \sqrt{\hbar c/e B} = 1$)
\begin{equation}
\phi_{m}({\bf r}) = (2^{m+1}\pi m!)^{-1/2} r^{m} e^{-im\phi}
 e^{-r^{2}/4} ,
\end{equation}
    where ${\bf r} = (x,y)$ is the position of the electron
     in the plane, $r = \sqrt{x^{2}+y^{2}}$ is the magnitude of
    ${\bf r}$, and $\phi = {\rm arctan}(y/x)$ is the angle between
    ${\bf r}$ and the $x$ axis. The ground state of the 2D electron gas
    at filling factor $\nu = 1$ is $|0\rangle = \prod_{m=0}^{\infty}
    e^{\dagger}_{m \uparrow}|{\rm vac}\rangle$, where $|{\rm vac}\rangle$
     is the empty conduction band, and $e^{\dagger}_{m \uparrow}$
    creates a spin-$\uparrow$ electron in the state $\phi_{m}({\bf r})$.
     Thus the 2D electron gas at filling factor $\nu = 1$ is a
        ferromagnet, with all electron spins aligned parallel to
     the magnetic field\fnm{a}\fnt{a}{The spins are parallel to
     the magnetic field---instead of antiparallel, as for free
    electrons---because the Land\'{e} $g$ factor is negative in GaAs.}.
    The ground state of the 2D electron gas at higher odd-integer
     filling factors ($\nu = 3,5,\ldots$) is also a ferromagnet,
    in which the lowest $\nu$ spin-$\uparrow$ Landau levels and
     the lowest $(\nu - 1)$ spin-$\downarrow$ Landau levels are
  completely filled, and all other Landau levels are empty. In
   the following we study the elementary charged excitations of
    a 2D electron gas at $\nu = 1$. With a few modifications, the
    results also apply to the higher odd-integer filling
     factors $\nu = 3,5,\ldots$.

\subsection{Elementary charged excitations at $\nu = 1$}
  \noindent The elementary charged excitations of a 2D electron
   gas at $\nu = 1$ are the excitations obtained by adding or
  removing an electron from the state $|0\rangle$. One must
 keep in mind that at filling factor $\nu = 1$ adding an electron
  is equivalent to {\em removing} a flux quantum $\Phi_{0} =
 h c/e$, while removing an electron is equivalent to {\em adding}
  a flux quantum\citeup{Laughlin}. The elementary charged excitations
  are eigenstates of the Hamiltonian $H = Z + V^{ee}$, where
\begin{equation}
\label{eq:zeeman}
 Z = \frac{1}{2} |g_{e}| \mu_{B} B  \sum_{m} (
       e^{\dagger}_{m \downarrow} e_{m \downarrow} +
       e_{m \uparrow} e^{\dagger}_{m \uparrow} ) ,
\end{equation}
\begin{equation}
\label{eq:vee}
V^{ee} = \frac{1}{2}  \sum_{\sigma \sigma'} \sum_{m m' m'' m'''}
 V^{ee}_{m m' m'' m'''} e^{\dagger}_{m \sigma}
    e^{\dagger}_{m' \sigma'} e_{m'' \sigma'} e_{m''' \sigma} .
\end{equation}
 Here $\sigma$ is the spin of an electron ($\sigma =$ $\uparrow$ or
   $\downarrow$), and $V^{ee}_{m m' m'' m'''}$ is a matrix element of
  the electron-electron interaction $V^{ee}(r) = e^{2}/\epsilon r$,
   where $\epsilon$ is the background dielectric constant. We have taken
   the energy of the state $|0\rangle$ as the zero of our energy
   scale. The nature of the elementary charged excitations is determined
    by the interplay between the Zeeman energy $Z$ and the
   electron-electron interaction $V^{ee}$. It is convenient to introduce
   the parameter $g = \frac{1}{2}|g_{e}|\mu_{B}B/(e^{2}/\epsilon \ell)$,
   which is the ratio of the Zeeman energy $\frac{1}{2}|g_{e}|\mu_{B}B$
   to the typical Coulomb energy $e^{2}/\epsilon \ell$. For
  noninteracting electrons ($g = \infty$) the elementary charged
   excitations are spin-$\downarrow$ electrons
   ($e^{\dagger}_{m \downarrow}|0\rangle$) and spin-$\uparrow$ holes
   ($e_{m \uparrow}|0\rangle$). For large but finite $g$ the
   electron-electron interaction can be treated using perturbation
    theory. Because $V^{ee}$ contains no terms that flip the spin
   of an electron, the elementary charged excitations at large $g$
    still have spin $\frac{1}{2}$. However, as $g$ is lowered the
   nature of the elementary charged excitations changes
    dramatically---they become Skyrmions.

   Skyrmions are twists in the spin density of the
  2D electron gas. Figure~1 shows a slice of the spin profile
   of a Skyrmion along the $x$ axis.
   The full spin profile is obtained by rotating the
    slice around a vertical axis through $x = 0$ (the $z$ axis).
    Figures~1(a) and 1(b) show the spin profile of a negatively
   and a positively charged Skyrmion, respectively.
  The spin points down at the center of the Skyrmion ($x = 0$).
   As one moves radially outward the spin rotates smoothly until
   it points up at the edge of the Skyrmion ($x = \pm \lambda$).
    To understand the connection between the sense of the spin
    twist and the charge of the Skyrmion we trace the spin
   along a circular path lying in a plane perpendicular to
     the paper (the $x$-$y$ plane) that encloses the
     center of the Skyrmion. For the negatively
    charged Skyrmion shown in Fig.~1(a) the spin
   precesses as if a magnetic flux $-\Phi_{0}$ had been inserted
    at the origin. Since at $\nu = 1$ removing a flux quantum
     is equivalent to adding an electron, the spin twist in
     Fig.~1(a) has charge $-e$. For the positively charged
    Skyrmion shown in Fig.~1(b) the spin precesses in the
   opposite sense yielding an excitation with charge $+e$.

\subsection{Variational approach to Skyrmions}
   \noindent Because Skyrmions have a spin much larger than
   $\frac{1}{2}$ one cannot calculate their energy by starting
  from the noninteracting case and treating $V^{ee}$ using
   perturbation theory. A similar situation arises in
   superconductivity, where the energy of the superconducting state
   cannot be obtained from the normal state by treating
   the pairing interaction between the electrons using perturbation
    theory. The BCS treatment of the superconducting state provides
   a clue for the description of Skyrmions. The variational states
   proposed by Fertig, Brey, C\^{o}t\'{e}, and MacDonald\citeup{Fertig}
     are
\begin{eqnarray}
\label{eq:psi-}
|\psi_{-}\rangle & = &  \prod_{m=0}^{\infty}
(u_{m} e^{\dagger}_{m \downarrow} e_{m+1 \uparrow} + v_{m})
e_{0 \uparrow}|0\rangle , \\
\label{eq:psi+}
|\psi_{+}\rangle & = &  \prod_{m=0}^{\infty}
(-u_{m} e^{\dagger}_{m+1 \downarrow} e_{m \uparrow} + v_{m})
e^{\dagger}_{0 \downarrow}|0\rangle .
\end{eqnarray}
   The state $|\psi_{-}\rangle$ ($|\psi_{+}\rangle$) describes a
   positively (negatively) charged Skyrmion localized at the origin
    of the $x$-$y$ plane\fnm{b}\fnt{b}{The subscript $\pm$ on
    $|\psi_{\pm}\rangle$ denotes the sign of the topological
   charge $Q = \pm 1$. The electric charge is given by $-Q e$.}.
     The state $|\psi_{-}\rangle$ ($|\psi_{+}\rangle$) is analogous
      to the BCS state\citeup{Schrieffer},
     except that the pairing now occurs between a spin-$\downarrow$
     electron (spin-$\uparrow$ hole) in the $m$\/th angular
    momentum state and a spin-$\uparrow$ hole (spin-$\downarrow$
     electron) in the $(m+1)$th angular
     momentum state of the lowest Landau level.

    The energy of the Skyrmion is obtained by minimizing
    the expectation value of $H$ in the state $|\psi_{\pm}\rangle$,
    subject to the constraint $|u_{m}|^{2}+|v_{m}|^{2} = 1$.
    Figure~2 shows the energy and spin of a negatively charged
     Skyrmion as a function of the parameter $g$. The energy of a
      spin-$\frac{1}{2}$ quasielectron is also shown for comparison.
     A positively charged Skyrmion has the same energy (up to a
      $g$-independent constant) and spin as a negatively charged
     Skyrmion\citeup{Fertig}.
   The size of the stable spin twist is determined by a
    competition between the Zeeman energy, which opposes spin flips
   (and hence tries to minimize $\lambda$), and the exchange interaction,
    which favors a gradual spin reversal (and therefore tries to maximize
    $\lambda$). At large $g$ the energy cost of flipping spins is
   prohibitive and the lowest energy excitation is a spin-$\frac{1}{2}$
    quasiparticle. As $g$ is lowered the size of the Skyrmion increases
    and its energy falls below that of a spin-$\frac{1}{2}$ quasiparticle.
    Near $g = 0$ the Skyrmion becomes a macroscopic object whose size
    is determined by the sample boundary.

\subsection{Experimental evidence for Skyrmions}

  \noindent The first experimental evidence for large-spin charged
    excitations near $\nu = 1$ was obtained by Barrett, Dabbagh,
    Pfeiffer, West, and
  Tycko\citeup{Barrett} using an optically pumped NMR
    technique. By measuring the Knight shift of
     $^{71}$Ga nuclei embedded in a quantum well these authors
    were able to determine the degree of spin polarization of the 2D
     electron gas as a function of the filling
   factor $\nu$. If the charged excitations were spin-$\frac{1}{2}$
   quasiparticles one would expect the spin polarization to remain
    constant for $\nu < 1$, and decline gradually for $\nu > 1$.
   Instead, Barrett and coworkers observed a rapid decrease of the spin
    polarization on {\em both} sides of $\nu = 1$. This is consistent
     with positively and negatively charged excitations of large and
    equal spin---i.e.\ Skyrmions.

     Further evidence supporting the existence of Skyrmions at $\nu = 1$
   was provided by Schmeller, Eisenstein, Pfeiffer, and
        West\citeup{Schmeller}. They performed activated
      magnetotransport measurements of the energy gap $\Delta$ for
   creating a {\em pair} of quasiparticles at $\nu = 1$. The energy
    gap is found from the temperature dependence of the longitudinal
   resistance using $R_{xx} = R_{0} \exp(-\Delta/2 k_{B}T)$.
    The spin $s$ of the quasiparticle pair is obtained by
      measuring the change in $\Delta$ produced by tilting
    the total magnetic field $B_{\rm tot}$ away from the normal to the 2D
      plane (while keeping the perpendicular field $B_{\perp}$ constant).
     The energy gap is given by
\begin{equation}
\label{eq:gap}
\Delta = \Delta_{0}(B_{\perp}) + s |g_{e}|\mu_{B} B_{\rm tot} .
\end{equation}
     The first term, $\Delta_{0}(B_{\perp})$, is the contribution
      to the energy gap from $V^{ee}$. For an infinitely thin
    2D electron gas this term only depends on the perpendicular
    field controlling the orbital motion. The second term is the
    Zeeman energy of the quasiparticle pair. According to
     Eq.~(\ref{eq:gap}) the slope $\partial \Delta/ \partial B_{\rm tot}$
     at constant $B_{\perp}$ is just $s |g_{e}|\mu_{B}$.
     Schmeller and company measured the variation of $\Delta$
     with $B_{\rm tot}$ in GaAs quantum wells at several
      odd-integer filling factors.
    At $\nu = 3$ and $\nu = 5$ they observed slopes with $s = 1$,
    as expected for a pair of spin-$\frac{1}{2}$ quasiparticles.
    However, at $\nu = 1$ they observed a much larger slope with
     $s \approx 7$. This implies that the quasiparticles at
     $\nu = 1$ (but not those at $\nu = 3$ and $\nu = 5$) are
     Skyrmions with a spin of approximately 3.5.

     Evidence for Skyrmions at $\nu = 1$ was also obtained by Aifer,
      Goldberg and Broido\citeup{Aifer} using interband optical
     transmission spectroscopy. Their experiment consisted of a
    measurement of the absorption coefficients for right and left
   circularly polarized (RCP and LCP) light as a function of
      filling factor between $\nu = 0.6$ and $\nu = 1.4$.
     In a simple model\citeup{Aifer} the absorption coefficient for
     RCP (LCP) light is proportional to the number of available
    (i.e.\ unoccupied) states $N_{A_{\downarrow}}$ ($N_{A_{\uparrow}}$)
     in the spin-$\downarrow$ (spin-$\uparrow$) lowest Landau level.
     The number of available states, $N_{A_{\sigma}}$
       ($\sigma =$ $\downarrow$ or $\uparrow$), is related to the
      number of occupied states ($N_{\sigma}$) by $N_{A_{\sigma}} =
     N_{B}-N_{\sigma}$, where $N_{B}$ is the Landau level degeneracy.
      The spin polarization per electron is $S_{z} =
    (N_{\downarrow}-N_{\uparrow})/N$, where $N = N_{\downarrow} +
      N_{\uparrow}$ is the total number of electrons. The difference
      between the RCP and LCP absorption intensities thus
       provides a direct measurement of the degree of spin
     polarization of the electron gas. Using this technique,
      Aifer and coworkers found a rapid decay of the spin
     polarization on {\em both} sides of $\nu = 1$, which is
      inconsistent with quasiparticles of spin $\frac{1}{2}$.
       From the rate of decay of the spin polarization they deduced
     values of $2.5$ and $3.7$ for the Skyrmion spin in their samples.

\section{PL Spectroscopy of Quantum Hall Systems}
  \noindent Spectroscopic methods based on the radiative recombination
      of 2D electrons with photoexcited holes have developed into
      a powerful tool for studying the quantum Hall effect. Optical
       measurements can give information about fundamental aspects
     of quantum Hall systems. For example, photoluminescence (PL)
       and photoluminescence excitation (PLE) measurements yield
       transition energies, polarization studies are sensitive to
       the electron spin, and ultrafast pump-and-probe experiments
       monitor carrier relaxation processes on a sub-picosecond
      timescale. In this work we consider the PL spectroscopy of
       quantum Hall systems near $\nu = 1$.

\subsection{Intrinsic and extrinsic PL}
    \noindent PL studies of quantum Hall systems can be divided
     into two types according to the experimental setup employed.
       {\em Intrinsic} PL experiments study the recombination
     of 2D electrons with photoexcited holes in standard
      GaAs/Ga$_{1-x}$Al$_{x}$As heterojunctions and quantum
      wells\citeup{Turberfield,Heiman,Goldberg,Plentz}.
      Figure~3 shows the band diagram of a one-side
       modulation-doped GaAs/Ga$_{1-x}$Al$_{x}$As quantum well.
      The structure consists of a GaAs layer sandwiched
      between two Ga$_{1-x}$Al$_{x}$As barriers. Only the left
      Ga$_{1-x}$Al$_{x}$As barrier is doped, leading to an
       asymmetric band profile. The energy spectrum in the
       absence of a magnetic field consists of a series of
       conduction and valence subbands\citeup{Weisbuch}.
      Only the lowest conduction
       subband ($E_{0}$) and the highest valence
      subband ($HH_{0}$) are shown in Fig.~3, along with the
      corresponding $z$ wave functions. The electrons are
       confined near the interface between the GaAs layer
      and the left Ga$_{1-x}$Al$_{x}$As barrier, while
       the holes are confined on the opposite side of the
       well, near the interface between the GaAs layer and
       the right Ga$_{1-x}$Al$_{x}$As barrier. The dotted
       lines indicate the average positions of the electrons
       and holes along $z$. For simplicity we consider
      a model in which the electrons and holes are confined
      to infinitely thin planes separated by a distance $d$.
        This amounts to approximating the $z$ wave functions
        by $\delta$ functions. The effect of the finite extent
      of the $z$ wave functions on the PL spectrum is discussed
       in Sec.~4.7.

     {\em Extrinsic} PL experiments utilise specialized
      structures, in which a $\delta$-doped layer of acceptors
       is grown into the GaAs at a certain distance from the
       Ga$_{1-x}$Al$_{x}$As--GaAs
       interface\citeup{Buhmann,Joss,Kukushkin}. The recombination
      studied in extrinsic PL is between a 2D electron and a hole
       bound to an acceptor in the $\delta$-doped layer. Extrinsic
      PL is quite different from intrinsic PL because the acceptor
       is charge {\em neutral} in the initial state. The type of
      PL considered in this work is intrinsic PL.

      An intrinsic PL experiment consists of three stages.
       In the first stage electrons are promoted from the GaAs valence
       band into the GaAs conduction band by a pump laser tuned below
      the Ga$_{1-x}$Al$_{x}$As band gap. Pump intensities are kept low
     ($< 10^{-4}$ ${\rm W cm}^{-2}$) to prevent heating of the
       electron gas. In the second stage the photoexcited carriers
      relax to the lowest energy state within their respective energy
       bands. The intraband relaxation takes place on a timescale
      (picoseconds) which is short compared to the radiative
       recombination time (nanoseconds). In the third stage the
      photoexcited holes recombine with either photoexcited electrons
       or 2D electrons. Recording the number of emitted photons
      of frequency $\omega$ and polarization $\alpha$ ($\alpha = +1$
       for RCP and $\alpha = -1$ for LCP) yields the
       (polarization-resolved) PL spectrum $P_{\alpha}(\omega)$.

\subsection{PL spectrum}
       \noindent According to the Fermi Golden Rule, the PL spectrum
        is given by
\begin{equation}
\label{eq:fermit}
P_{\alpha}(\omega) = 2 \pi \sum_{i,f} n_{i}
 |\langle f|L_{\alpha}|i\rangle|^{2}
 \delta(\omega - E_{i} + E_{f}) .
\end{equation}
       Here $|i\rangle$ ($|f\rangle$) is the initial (final) state
       before (after) recombination, $n_{i}$ is the occupation of
      initial states, and $L_{\alpha}$ is the luminescence operator
       (given further below). The energy $\omega$ of the emitted photon
      is equal to the difference between the energy of the initial
       state ($E_{i}$) and
     the energy of the final state ($E_{f}$). Assuming complete
      relaxation of the photoexcited carriers prior to recombination,
      the occupation of initial states is $n_{i} = \frac{1}{Z}
      e^{-E_{i}/k_{B}T}$, where $Z = \sum_{i} e^{-E_{i}/k_{B}T}$ is
     the partition function at temperature $T$\@. PL studies of
      quantum Hall systems are done at temperatures as low as
     $T = 0.1$ K (corresponding to an energy of about $0.01$ meV).
     At these temperatures only the initial state with the lowest
     energy, $|i_{0}\rangle$, has significant occupation. Setting
       $n_{i} = \delta_{i,i_{0}}$ in Eq.~(\ref{eq:fermit}),
\begin{equation}
\label{eq:fermio}
P_{\alpha}(\omega) = 2 \pi \sum_{f}
 |\langle f|L_{\alpha}|i_{0}\rangle|^{2}
 \delta(\omega - E_{i_{0}} + E_{f}) ,
\end{equation}
     where $E_{i_{0}}$ is the energy of the state $|i_{0}\rangle$.
    As compared to the ground state $|0\rangle$, the initial state
     $|i_{0}\rangle$ has one more negatively charged excitation
     in the conduction band and a hole in the valence band.
       Provided the hole does not significantly perturb the
      electron gas, the negatively charged excitation is precisely
     the one studied in Sec.~3.1. We may therefore expect to see
      a signature of Skyrmions in the PL spectrum at $\nu = 1$.

\subsection{PL spectra near $\nu = 1$}
      \noindent A number of PL studies near $\nu = 1$ have
        been reported in the
        literature\citeup{Turberfield,Heiman,Goldberg,Plentz}.
       Below we summarize the results of two representative
        studies. Figure~4 shows the results of an experiment
       by Turberfield {\em et al.}\citeup{Turberfield} on a
      GaAs/Ga$_{0.68}$Al$_{0.32}$As single heterojunction.
       The figure shows
     the magnetic-field dependence of the energy of the $E_{0}$
     luminescence line arising from the recombination of
      2D electrons in the lowest conduction subband ($E_{0}$)
      with holes in the GaAs valence band. The inset shows the
     line shapes at selected magnetic fields near $\nu = 1$. The
      important features are:
\begin{romanlist}
\item There are two lines on the low-$B$ side of $\nu = 1$,
       and one line on the high-$B$ side.
\item The lower-energy line has an abrupt redshift
      as the magnetic field is increased past $\nu = 1$.
\end{romanlist}

     Polarization-resolved PL studies of GaAs/Ga$_{1-x}$Al$_{x}$As
     quantum wells and heterojunctions in the quantum Hall regime
      have been reported by Goldberg {\em et al.}\citeup{Goldberg}
     The PL spectra in wide quantum wells (with well widths of
      $400$--$500$ \mbox{\AA}) and single heterojunctions have
     the following features:
\begin{romanlist}
\item There is a RCP line and a LCP line on the low-$B$ side of
       $\nu = 1$, and only a LCP line on the high-$B$ side.
\item The LCP line has an abrupt redshift as the
       magnetic field is increased past $\nu = 1$.
\end{romanlist}

       The important features of the PL spectrum
        near $\nu = 1$ can be explained using a simple
       model\citeup{Muzykantskii,Cooper} in which the electrons
        occupy the lowest Landau level of the $E_{0}$ conduction
       subband, and the holes occupy the highest Landau level of
       the $HH_{0}$ valence subband (see Fig.~3). The electrons
       have spin $\uparrow$ or $\downarrow$, and the holes have
       $m_{j} = \pm \frac{3}{2}$. The lowest-energy transition
        in LCP is from the spin-$\uparrow$ electron state to the
       $m_{j} = +\frac{3}{2}$ hole state, and the lowest-energy
       transition in RCP is from the spin-$\downarrow$ electron
        state to the $m_{j} = -\frac{3}{2}$ hole state.

        In Fig.~5 we compare a PL experiment at $\nu = 1$ with
        a PL experiment at the filling factor just below $\nu = 1$
       where one spin-$\uparrow$ electron is missing from the lowest
       Landau level [Fig.~5(a)].
       We denote this filling factor by $\nu = 1^{-}$.
        At $\nu = 1$ the state before recombination
       has one extra spin-$\downarrow$ electron and a valence hole
        [Fig.~5(b)]. The photoexcited electron must have spin
        $\downarrow$ because the spin-$\uparrow$ Landau level is
       completely filled. At $\nu = 1^{-}$ the state before
       recombination has a filled spin-$\uparrow$ level and
       a valence hole. The photoexcited electron has filled up
        the hole in the spin-$\uparrow$ Landau level that was
        present in the ground state at $\nu = 1^{-}$.

       In the LCP transition a spin-$\uparrow$ electron recombines
       with a $m_{j} = +\frac{3}{2}$ hole. This transition can take
       place both at $\nu = 1$ and at $\nu = 1^{-}$ [Fig.~5(c)].
        We thus expect to see a LCP line on both sides of $\nu = 1$.
       In the RCP transition a spin-$\downarrow$ electron recombines
        with a $m_{j} = -\frac{3}{2}$ hole. This transition {\em can}
        take place at $\nu = 1$, but {\em not} at $\nu = 1^{-}$, where
       no spin-$\downarrow$ electrons are available in the initial state
       [Fig.~5(d)]. We thus expect to see a RCP line on the low-$B$
        side of $\nu = 1$, but not on the high-$B$ side.

       To explain the redshift of the LCP line we compare the
       states before and after LCP recombination at $\nu = 1$
       and $\nu = 1^{-}$ [Figs.~5(b) and (c)]. The spin-$\downarrow$
       electron and the valence hole in the initial state at
         $\nu = 1$ will bind to form an interband exciton. This
        means the initial state at $\nu = 1$ has a lower
       energy than the initial state at $\nu = 1^{-}$, which
       has an unbound valence hole. The energy difference is
         $\Delta E_{i} = -B_{X}$, where $B_{X}$ is the binding
        energy of the interband exciton\citeup{Cooper}. Similarly,
       the spin-$\downarrow$ electron and the spin-$\uparrow$ hole
        in the final state at $\nu = 1$ will bind to form a spin
       wave. Hence the final state at $\nu = 1$
       has a lower energy than the final state at $\nu = 1^{-}$,
       which has an unbound spin-$\uparrow$ hole. The energy
        difference between the final states is $\Delta E_{f} = -B_{SW}$,
        where $B_{SW}$ is the binding energy of the spin
       wave\citeup{Cooper}. The shift of the LCP luminescence
       line is given by $\Delta \omega = \Delta E_{i} - \Delta E_{f}
        = B_{SW} - B_{X}$. Because the spin-$\downarrow$ electron
         and the valence hole are spatially separated along $z$,
        while the spin-$\downarrow$ electron and the spin-$\uparrow$
        hole are both in the same plane, the spin wave has a larger
         binding energy than the interband exciton. Hence the LCP
       line shifts to {\em lower} energies (redshift) as the
        magnetic field is increased past $\nu = 1$.

\section{Optics with Skyrmions in Disordered Systems}
      \noindent The simple model discussed in Sec.~3.3
       is based on a quasiparticle picture of the
        charged excitations at $\nu = 1$. What if the charged
       excitations were Skyrmions, instead of spin-$\frac{1}{2}$
        quasiparticles? In what way does the recombination
       spectrum of Skyrmion differ from that of a
        spin-$\frac{1}{2}$ quasielectron? These are the
        questions this section aims to address.

       Besides being of interest for the PL spectrum
        at $\nu = 1$, these questions are also of broader
      relevance to the optical properties of quantum Hall
       systems at other filling factors. The $\nu = 1$
      state serves as a prototype for spin-polarized states
       at higher odd-integer filling factors
      ($\nu = 3,5,\ldots$), and, via the composite-fermion
     picture\citeup{Jain}, also for spin-polarized states
      at fractional filling factors ($\nu = \frac{1}{3},
     \frac{2}{5},\ldots$). Indeed, it has been
        suggested\citeup{Kamilla} that the elementary
     charged excitations at $\nu = \frac{1}{3},\frac{2}{5},\ldots$
      can be regarded as Skyrmions of composite fermions. An
         understanding of the recombination spectrum
      at $\nu = 1$ could therefore provide important clues
       about the recombination spectrum of spin-polarized
      states at other filling factors.

      We start our investigation of the optical properties
       of Skyrmions with a study of disordered systems. The
      localization of the photoexcited hole by the disorder
      potential considerably simplifies the mathematical
       treatment of the initial state. The optical properties
       of Skyrmions in pure systems are discussed
       in Sec.~5.

\subsection{Sources of disorder}
       \noindent Sources of disorder in a modulation-doped
         quantum well of the type shown in Fig.~3 include:
\begin{itemlist}
\item Ionized impurities in the doped layer.
\item Residual impurities in the undoped layers.
\item Roughness of the interface between the
      GaAs and Ga$_{1-x}$Al$_{x}$As layers.
\item Fluctuations in the alloy composition of
      the Ga$_{1-x}$Al$_{x}$As layers.
\end{itemlist}
       To model the disorder we introduce disorder potentials
       $V^{e}({\bf r})$ and $V^{h}({\bf r})$ acting on the
      electrons and holes, respectively.
       The dominant contribution to $V^{e}({\bf r})$ comes
      from ionized impurities in the doped layer. To achieve
       a higher electron mobility an undoped spacer layer is
       grown between the doped layer and the GaAs well
       (see Fig.~3). The ability to spatially separate the
       electrons from the donor impurities accounts for the
        extremely high mobilities achieved
       in modulation-doped heterostructures when compared
       to bulk semiconductors. The larger separation between
       the doped layer and the holes reduces the contribution
      of this type of disorder to $V^{h}({\bf r})$.

       Disorder due to interface roughness is expected to
        dominate $V^{h}({\bf r})$. The holes are confined near
       the so-called ``inverted'' interface (i.e.\ GaAs grown on
      top of Ga$_{1-x}$Al$_{x}$As) which is known\citeup{Inoue}
        to have a higher
       degree of disorder than the ``normal'' interface
        (Ga$_{1-x}$Al$_{x}$As on top of GaAs).
       It was, in fact, the difficulty of growing a smooth inverted
       interface that prompted the introduction of the one-side
        modulation-doped quantum well\citeup{Meynadier}. Doping
        only on the side of the normal interface pulls the
       electrons away from the inverted interface\citeup{Cho},
       thus significantly reducing the contribution of this
         type of disorder to $V^{e}({\bf r})$.

       In the following we neglect $V^{e}({\bf r})$ compared
        to $V^{h}({\bf r})$. This is reasonable given the very
      high mobility of the samples used in recent
        experiments\citeup{Turberfield,Goldberg}. Note that
         $V^{h}({\bf r})$ does not affect the mobility because
       no holes are present prior to illumination.

\subsection{Initial state in a disordered system}
     \noindent Consider the optical recombination of a Skyrmion
   with a hole localized near a minimum in $V^{h}({\bf r})$. We choose
      the location of the minimum as the origin of our coordinate
      system. Provided the disorder potential varies slowly on
     the scale of $\ell$, the hole is in the
      $m = 0$ state. The Skyrmion also becomes localized near
     the origin due to the Coulomb attraction between the Skyrmion
     and the hole. Assuming complete energy relaxation prior
      to recombination, the initial state is given by
\begin{equation}
\label{eq:istate}
|i_{0}\rangle = \prod_{m=0}^{\infty}
(-u_{m} e^{\dagger}_{m+1 \downarrow} + v_{m}e^{\dagger}_{m \uparrow})
e^{\dagger}_{0 \downarrow} h^{\dagger}_{0 m_{j}} |{\rm vac}\rangle .
\end{equation}
    The parameters $u_{m}$ and $v_{m}$ are determined by
     minimizing the expectation value of
    $H = \varepsilon^{h}_{0m_{j}} + \,Z\, + \,V^{ee} + \,V^{eh}$
     in the state $|i_{0}\rangle$, subject to the constraint
      $|u_{m}|^{2} + |v_{m}|^{2} = 1$. Here
\begin{equation}
\label{eq:ehole}
\varepsilon^{h}_{0m_{j}} = E_{g} + \frac{e B}{2 \mu c} +
m_{j} g_{h} \mu_{B} B
\end{equation}
    is the energy of the hole\fnm{c}\fnt{c}{Recall that we
    have chosen the energy of the state $|0\rangle$ as the zero
     of our energy scale.}, where $E_{g}$ is the bandgap at
     zero magnetic field, $\mu = m_{e}m_{h}/(m_{e}+m_{h})$ is
     the reduced mass, and $g_{h}$ is the Land\'{e} $g$-factor for
      the hole. $Z$ and $V^{ee}$ are
     given by Eqs.~(\ref{eq:zeeman}) and (\ref{eq:vee}),
    respectively, and
\begin{equation}
\label{eq:veh}
V^{eh} = \sum_{\sigma,m_{j}} \sum_{m m' m'' m'''}
 V^{eh}_{m m' m'' m'''} e^{\dagger}_{m \sigma}
    h^{\dagger}_{m' m_{j}} h_{m'' m_{j}} e_{m''' \sigma}
\end{equation}
     is the electron-hole interaction. Here $V^{eh}_{m m' m'' m'''}$
    is a matrix element of $V^{eh}({\bf r}) = e^{2}/\epsilon |{\bf r} +
     \hat{z}d|$, where $d$ is the separation between the electron
    and hole planes along $z$ (see Fig.~3).

      To evaluate the expectation value we use the relations
\begin{eqnarray}
\label{eq:downdown}
\langle i_{0}|e^{\dagger}_{m \downarrow} e_{m \downarrow}|
        i_{0} \rangle & = & |u_{m-1}|^{2} ,\\
\label{eq:upup}
\langle i_{0}|e^{\dagger}_{m \uparrow} e_{m \uparrow}|
        i_{0} \rangle & = & |v_{m}|^{2} ,\\
\label{eq:downup}
\langle i_{0}|e^{\dagger}_{m+1 \downarrow} e_{m \uparrow}|
        i_{0} \rangle & = & -u^{*}_{m}v_{m} .
\end{eqnarray}
      The first (second) relation gives the probability of finding
      an electron with spin $\downarrow$ ($\uparrow$) in the $m$\/th
        angular momentum state. We define $u_{-1} \equiv -1$ and
      $v_{-1} \equiv 0$ to account for the definite occupation of
       the $m = 0$ state by a spin-$\downarrow$ electron.
       The third relation expresses the pairing correlation between
     a spin-$\downarrow$ electron in the $(m+1)$\/th angular momentum
      state and a spin-$\uparrow$ electron in the $m$\/th angular
      momentum state. With the aid of these relations we find
\begin{equation}
\label{eq:sandwich}
\langle i_{0}|H|i_{0} \rangle = \varepsilon^{h}_{0 m_{j}} +
U^{Z} + U^{H} - U^{ex} - U^{sk} - U^{eh} ,
\end{equation}
      where
\begin{eqnarray}
\label{eq:uz}
U^{Z} & = & \frac{1}{2}|g_{e}|\mu_{B}B \sum_{m=0}^{\infty}
(|u_{m-1}|^{2} - |v_{m}|^{2} + 1) , \\
\label{eq:uh}
U^{H} & = & \frac{1}{2} \sum_{m,m'=0}^{\infty} V^{ee}_{m m'm' m}
(|u_{m-1}|^{2} + |v_{m}|^{2} - 1)(|u_{m'-1}|^{2} + |v_{m'}|^{2} - 1) ,\\
\label{eq:uex}
U^{ex} & = & \frac{1}{2} \sum_{m, m'=0}^{\infty} V^{ee}_{m m'm m'}
(|u_{m-1}|^{2}|u_{m'-1}|^{2} + |v_{m}|^{2}|v_{m'}|^{2} - 1) ,\\
\label{eq:usk}
U^{sk} & = & \sum_{m, m'=0}^{\infty}
V^{ee}_{m m' m-1 m'+1} u^{*}_{m-1} v_{m-1} u_{m'} v^{*}_{m'}, \\
\label{eq:ueh}
U^{eh} & = & \sum_{m=0}^{\infty} V^{eh}_{m 0 0 m}
(|u_{m-1}|^{2} + |v_{m}|^{2} - 1).
\end{eqnarray}

     The expectation value of $H$ is the sum of six terms.
      The first term is the energy of the hole. The second
     term ($U^{Z}$) is the Zeeman energy of the Skyrmion.
      Since $u_{m} \rightarrow 0$ and $v_{m} \rightarrow 1$
      when $m \rightarrow \infty$ the sum over $m$ converges.
     The third term ($U^{H}$) has contributions from the direct
     Coulomb interaction among the electrons (Hartree energy),
     the Coulomb interaction between the electrons and the
     positive background, and the Coulomb energy of the positive
     background itself. The fourth term ($U^{ex}$) is exchange
      interaction between electrons of parallel spin.
     The factor of $-1$ inside the parentheses comes from
    subtracting the exchange energy of the state $|0\rangle$.
       The fifth term ($U^{sk}$) is the correlation energy which
     arises from the pairing of a spin-$\downarrow$ electron in
    the $(m+1)$\/th angular momentum state with a spin-$\uparrow$
     electron in the $m$\/th angular momentum state. Finally,
     the sixth term ($U^{eh}$) is the Coulomb interaction between
      the electrons and the hole.

      The results of a numerical minimization of
       $\langle i_{0}|H|i_{0} \rangle$ are shown in Fig.~6.
       Figure~6(a) shows the difference between the
        minimized energy of the state $|i_{0}\rangle$ (consisting
       of a Skyrmion and a hole) and the energy of the state
      $e^{\dagger}_{0\downarrow} h^{\dagger}_{0m_{j}} |0\rangle$
       (consisting of an spin-$\downarrow$ electron and a hole
        against a $\nu = 1$ background). Figure~6(b) shows the
      corresponding difference in spin between the two states.
        The energy and spin of the Skyrmion are determined by
       three competing effects. As in Fig.~2 the Skyrmion shrinks
       with increasing Zeeman coupling. The difference between the
       state $|i_{0}\rangle$ and the state $|\psi_{+}\rangle$ is
       the presence of the photoexcited hole. The electron-hole
        interaction favors a charge distribution for the electrons
       that is strongly peaked in the vicinity of the hole and
       therefore Skyrmions of small size. Bringing the hole
       closer to the electron plane increases the strength
       of the electron-hole interaction, thereby reducing the
        Skyrmion size. Upon decreasing $g$ at a fixed value
        of $d$ there is a threshold to Skyrmion formation.
       When the hole is far away from the electron plane
       ($d = \infty$) the threshold value of $g$ is
        approximately $0.025$. Reducing $d$ lowers the
        value of $g$ below which Skyrmions form. When the
        hole comes too close to the electron plane ($d < \ell$)
         there is no Skyrmion formation at any value of $g$.
       Thus we may expect to see an optical signature of
        Skyrmions in wide ($d > \ell$) quantum wells,
          but not in narrow ($d < \ell$) wells.

\subsection{Moments of the PL spectrum}
     \noindent We treat the optical recombination in the
      electric-dipole approximation. The luminescence operators for
       RCP and LCP transitions are given by
\begin{equation}
\label{eq:lplus}
L_{+} = \mu_{+} \sum_{m} e_{m\downarrow} h_{m,-3/2} ,
\end{equation}
\begin{equation}
\label{eq:lminus}
L_{-} = \mu_{-} \sum_{m} e_{m\uparrow} h_{m,3/2} ,
\end{equation}
           respectively, where $\mu_{\pm}$ is the product of
         the interband dipole matrix element and the overlap
          between the electron and hole $z$ wave functions.
         There is an important difference between the optical
        recombination of a spin-$\frac{1}{2}$ quasielectron
         and the optical recombination of a Skyrmion.
          While recombination of a spin-$\frac{1}{2}$
         quasielectron leaves either no spin flips (in the
         case of RCP) or one spin flip (in the case of LCP)
         in the final state, recombination of a Skyrmion
         leaves a large number of spin flips in the final state.
          For a Skyrmion with spin $S_{z}$, the number of spin
          flips left is $|S_{z} + \frac{1}{2}|$
          in the case of RCP, and $|S_{z} - \frac{1}{2}|$
        in the case of LCP.

           The PL spectrum is given by Eq.~(\ref{eq:fermio}).
          The detailed line shape is difficult to calculate
          owing to the complicated nature of the final states
           containing multiple spin flips. However, the existing
           PL studies discussed in Sec.~3.3 suggest
             that the quantity of primary interest is the
           luminescence energy
\begin{equation}
 \langle \omega_{\alpha} \rangle = \int d \omega\,\omega
      P_{\alpha}(\omega) .
\end{equation}
          The usefulness of moments of the PL
             spectrum was demonstrated by Apalkov and
            Rashba\citeup{Apalkov} in the context of the
            fractional quantum Hall effect.
           The luminescence energy (and other moments of
            the spectrum) can be obtained directly from the
           initial state $|i_{0}\rangle$. Integrating
          Eq.~(\ref{eq:fermio}) with respect to $\omega$ we obtain
\begin{equation}
 \langle \omega_{\alpha} \rangle = 2 \pi \sum_{f}
(E_{i_{0}}-E_{f}) |\langle f|L_{\alpha}|i_{0}\rangle |^{2} .
\end{equation}
            If $|i_{0}\rangle$ and $|f\rangle$ are eigenstates of $H$
         (i.e.\ $H|i_{0}\rangle = E_{i_{0}}|i_{0}\rangle$ and
           $H|f\rangle = E_{f}|f\rangle$) this can be written as
\begin{equation}
 \langle \omega_{\alpha} \rangle = 2 \pi \sum_{f}
\langle i_{0}|L^{\dagger}_{\alpha}|f\rangle
\langle f|[L_{\alpha},H]|i_{0}\rangle ,
\end{equation}
           where $[L_{\alpha},H] = L_{\alpha}H-HL_{\alpha}$.
          We now sum over all final states within the lowest Landau
           level. This yields the projection operator
           $P = \sum_{f}|f\rangle \langle f|$ onto the lowest
          Landau level. Since $L_{\alpha}$, $H$, and $|i_{0}\rangle$
           are already in the lowest Landau level, the projection
           operator acts as the unit operator. Hence
\begin{equation}
 \langle \omega_{\alpha} \rangle = 2 \pi
\langle i_{0}|L^{\dagger}_{\alpha}[L_{\alpha},H]|i_{0}\rangle .
\end{equation}
           Expressions for the other moments can be derived in a
            similar way. The $n$\/th-order moment
            $\langle \omega^{n}_{\alpha} \rangle =
           \int d\omega \,\omega^{n} P_{\alpha}(\omega)$ is given by
\begin{equation}
\label{eq:moments}
\langle \omega^{n}_{\alpha} \rangle = 2 \pi
\langle i_{0}|L^{\dagger}_{\alpha}[L_{\alpha},H]_{n}|i_{0}\rangle ,
\end{equation}
           where $[L_{\alpha},H]_{n}$ is defined by
          $[L_{\alpha},H]_{n} = [[L_{\alpha},H]_{n-1},H]$ and
          $[L_{\alpha},H]_{0} = L_{\alpha}$. It is convenient to
          redefine $\langle \omega^{n}_{\alpha} \rangle =
       \langle \omega^{n}_{\alpha} \rangle /
       \langle \omega^{0}_{\alpha} \rangle$, where
      $\langle \omega^{0}_{\alpha} \rangle$ is the integrated
          intensity.

\subsection{PL energies}
     \noindent To calculate the energies of the RCP and LCP lines we
     use Eq.~(\ref{eq:moments}) with $n = 0,1$ and the initial
      state given by Eq.~(\ref{eq:istate}). Working out the
      commutator and applying Eqs.~(\ref{eq:downdown})--(\ref{eq:downup}),
      we find
\begin{eqnarray}
\label{eq:wplus}
 \langle \omega_{+} \rangle & = &
\varepsilon^{h}_{0,-3/2} +
\varepsilon^{e}_{0 \downarrow} - U^{eh}_{0 \downarrow} , \\
\label{eq:wminus}
\langle \omega_{-} \rangle & = &
\varepsilon^{h}_{0,3/2} +
 E^{e}_{0 \uparrow} - U^{eh}_{0 \uparrow} ,
\end{eqnarray}
      where
\begin{eqnarray}
\label{eq:uehup}
U^{eh}_{0 \uparrow} & = & V^{eh}_{0000} +
\sum_{m=2}^{\infty} V^{eh}_{m00m} |u_{m-1}|^{2} +
\sum_{m=1}^{\infty} V^{eh}_{m00m} |v_{m}|^{2} , \\
\label{eq:uehdown}
U^{eh}_{0 \downarrow} & = &
\sum_{m=1}^{\infty} V^{eh}_{m00m} |u_{m-1}|^{2} +
\sum_{m=0}^{\infty} V^{eh}_{m00m} |v_{m}|^{2} ,
\end{eqnarray}
      and
\begin{equation}
\label{eq:eup}
E^{e}_{0 \uparrow} =
\frac{\varepsilon^{e}_{1 \downarrow} +
      \varepsilon^{e}_{0 \uparrow}}{2} - \sqrt{ \left(
\frac{\varepsilon^{e}_{1 \downarrow} -
      \varepsilon^{e}_{0 \uparrow}}{2}
\right)^{2} + |U^{sk}_{0}|^{2}} ,
\end{equation}
       with
\begin{eqnarray}
\label{eq:emup}
\varepsilon^{e}_{m \uparrow} & = &  -\frac{1}{2}|g_{e}|\mu_{B}B +
 U^{H}_{m} - U^{ex}_{m \uparrow} - V^{eh}_{m00m} , \\
\label{eq:emdown}
\varepsilon^{e}_{m \downarrow} & = & \frac{1}{2}|g_{e}|\mu_{B}B +
 U^{H}_{m} - U^{ex}_{m \downarrow} - V^{eh}_{m00m} ,
\end{eqnarray}
      and
\begin{eqnarray}
\label{eq:uhm}
U^{H}_{m} & = & \sum_{m'=0}^{\infty}
V^{ee}_{mm'm'm} (|u_{m'-1}|^{2} + |v_{m'}|^{2} - 1) , \\
\label{eq:uexmup}
U^{ex}_{m \uparrow} & = &
\sum_{m'=0}^{\infty} V^{ee}_{mm'mm'} |v_{m'}|^{2} , \\
\label{eq:uexmdown}
U^{ex}_{m \downarrow} & = &
\sum_{m'=0}^{\infty} V^{ee}_{mm'mm'} |u_{m'-1}|^{2} , \\
\label{eq:uskm}
U^{sk}_{m} & = &
\sum_{m'=0}^{\infty} V^{ee}_{mm'+1\,m+1\,m'} u^{*}_{m'}v_{m'} .
\end{eqnarray}

     The RCP luminescence energy given by Eq.~(\ref{eq:wplus}) is
      the sum of three terms. The first term ($\varepsilon^{h}_{0,-3/2}$)
      is the energy of the hole. The second term
        ($\varepsilon^{e}_{0 \downarrow}$) is the energy of the
        spin-$\downarrow$ electron at the center of the Skyrmion.
      The third term ($U^{eh}_{0 \downarrow}$) is the Coulomb
       interaction between the remaining electrons (i.e.\ all
      those {\em except} the spin-$\downarrow$ electron at the center)
       and the valence hole. The LCP luminescence energy
       given by Eq.~(\ref{eq:wminus})
       also has three terms, of which only the second term
     ($E^{e}_{0 \uparrow}$) needs further explanation.
      This term represents the energy of the spin-$\uparrow$
       electron at the the center of the Skyrmion.
      Removing a spin-$\uparrow$ electron from the $m = 0$ state
      breaks the pair bond with the spin-$\downarrow$ electron
      in the $m = 1$ state. Hence $E^{e}_{0 \uparrow}$ involves
       not only $\varepsilon^{e}_{0 \uparrow}$, but also
      $\varepsilon^{e}_{1 \downarrow}$ and the pair interaction
       $U^{sk}_{0}$.

     In Fig.~7 we show the mean luminescence energy
\begin{equation}
 \langle \omega \rangle = \frac{\langle \omega_{+} \rangle
 + \langle \omega_{-} \rangle}{2}
\end{equation}
        as a function of $g$ and $d$.
         Starting from $g = 0$, the mean luminescence energy
         increases monotonically until $g$ reaches a threshold $g_{c}$.
       Above the threshold $\langle \omega \rangle$ does not vary
        with $g$. The threshold is $g_{c} = 0.025$ when $d = \infty$
        and becomes smaller as the hole is brought closer
       to the electron plane. By setting $u_{m} = 0$ and $v_{m} = 1$
       in Eqs.~(\ref{eq:wplus})--(\ref{eq:uskm}) (but keeping
        $u_{-1} = -1$) we can show that for a spin-$\frac{1}{2}$
        quasielectron $\langle \omega \rangle$ does {\em not} depend
       on $g$. Hence a variation of $\langle \omega \rangle$ with
       $g$ signals the presence of Skyrmions in the initial state.

\subsection{Redshift}
       \noindent We now consider the PL spectrum
         at filling factor $\nu = 1^{-}$.
       In the context of the present model---which includes disorder
       for the hole---the initial state at $\nu = 1^{-}$ is given by
       $|i_{0}\rangle = h^{\dagger}_{0 m_{j}}|0\rangle$. The energy
        of this state is
\begin{equation}
E_{i_{0}} = \varepsilon^{h}_{0 m_{j}} -
 \sum_{m=0}^{\infty} V^{eh}_{m00m} .
\end{equation}
       Since there are no spin-$\downarrow$ electrons in the initial
       state the RCP line is missing from the PL spectrum. The final
       state after LCP recombination is $|f\rangle = e_{0 \uparrow}
       |0\rangle$ with energy
\begin{equation}
E_{f} = \frac{1}{2}|g_{e}|\mu_{B}B +
 \sum_{m=0}^{\infty} V^{ee}_{m0m0} .
\end{equation}
      Thus the PL spectrum at $\nu = 1^{-}$ consists of a single
      LCP line at
\begin{equation}
\label{eq:ominus}
\omega_{-} = \varepsilon^{h}_{0,3/2}
 - \frac{1}{2}|g_{e}|\mu_{B}B
 - \sum_{m=0}^{\infty} V^{ee}_{m0m0}
 - \sum_{m=0}^{\infty} V^{eh}_{m00m} .
\end{equation}

       The redshift of the LCP line is given by the difference
        between Eq.~(\ref{eq:wminus}) and Eq.~(\ref{eq:ominus}).
       For a spin-$\frac{1}{2}$ quasielectron the redshift
       {\em does not} depend on $g$. Setting $u_{m} = 0$ and
       $v_{m} = 1$ we find
\begin{equation}
\Delta \omega_{-} = V^{ee}_{0000} - V^{eh}_{0000} .
\end{equation}
         For a Skyrmion the redshift {\em does} depend on $g$.
        Hence a measurement of the $g$ dependence of the
          redshift provides a way to distinguish
         between the optical recombination of a Skyrmion and
         the optical recombination of a spin-$\frac{1}{2}$
          quasielectron.

       In practice $g$ may be varied by tilting the magnetic
        field away from the normal to the 2D plane or by applying
      hydrostatic pressure. The tilted-field method was
         discussed earlier in Sec.~2.3. While the
         spin couples to the total field ($B_{\rm tot}$), the orbital
     motion is controlled by the perpendicular field ($B_{\perp}$).
      Hence $g = g_{\perp}/\cos \theta$, where $g_{\perp} =
       \frac{1}{2}|g_{e}|\mu_{B}B_{\perp}/(e^{2}/\epsilon \ell_{\perp})$
      (with $\ell_{\perp} = \sqrt{\hbar c/e B_{\perp}}$), and $\theta$
       is the tilting angle. Note that tilting the
        magnetic field increases $g$. An alternative way of
      varying $g$ is by applying hydrostatic
       pressure\citeup{Maude,Holmes}.
       The Land\'{e} $g$ factor in GaAs ($g_{e} = -0.44$) differs
        from the $g$ factor for free electrons ($g_{e} = 2.0$) due
      to the mixing between the conduction band and the spin-orbit
       split valence band. Applying hydrostatic pressure increases
        the bandgap, thereby reducing the amount of mixing between
        the two bands. As a result the $g$ factor becomes more
        free-electron like when pressure is applied. The variation
      of $g_{e}$ with applied pressure has been calculated by Holmes
        {\em et al.}\citeup{Holmes} using ${\bf k} \cdot {\bf p}$
      perturbation theory. In the pressure range from $0$ to $10$
        kbar they find $g_{e} = -0.43 + 0.02 p$, where $p$ is the
       pressure (in kbar). Note that $|g_{e}|$ {\em decreases} upon
         applying pressure.

        In Fig.~8 we show the calculated variation
         of the redshift with tilting angle ($\theta$) and applied
       pressure ($p$), for a GaAs/Ga$_{1-x}$Al$_{x}$As quantum well
      with $n_{s} = 10^{11}$~cm$^{-2}$ and a separation
         of $400$~\mbox{\AA}
        between the electron and hole planes. Upon tilting the field
        by $60^{\rm o}$ the red shift increases by $0.6$ meV, while
       raising the pressure from $1$ to $10$ kbar reduces the
      red shift by $0.4$ meV. In both cases the change of the redshift
       is well within the experimental resolution (typically
        $0.1$ meV).

\subsection{Finite-size effect}
       \noindent Thus far we have considered a model in which
      the electrons and the hole are confined to infinitely thin planes
         separated by a distance $d$. In reality the electron and
        hole wave functions have a finite extent along $z$ (Fig.~3).
         We now discuss the effect of the finite extent of the
          $z$ wave functions on the PL spectrum for two particular
         device designs used in recent experiments: an asymmetric
         quantum well, and a single heterojunction.

         We first consider the tilted-field experiment for an
        asymmetric quantum well. The parallel component of the
         magnetic field ($B_{\parallel}$) causes a diamagnetic
        shift of the electron and hole subband energies ($E_{0}$
          and $HH_{0}$). This leads to an additional
         $\theta$ dependence of the PL energy, which does not
        occur in the strictly 2D system. For a weak parallel field
         the diamagnetic shift may be calculated using perturbation
        theory\citeup{Ando}. The additional $\theta$ dependence of
         the PL energy is then given by
\begin{equation}
\label{eq:deform}
\Delta \langle \omega_{\alpha} \rangle =
\frac{\omega_{ce} (\Delta z_{e})^{2} +
      \omega_{ch} (\Delta z_{h})^{2} }{2 \ell_{\perp}^{2}}
\tan^{2} \theta ,
\end{equation}
        where $\Delta z_{e}$ ($\Delta z_{h}$) is the spread of
         the electron (hole) wave function along $z$, and
        $\omega_{ce}$ ($\omega_{ch}$) is the electron (hole)
         cyclotron energy. For example, if $\Delta z_{e} = 50$
        \mbox{\AA} and $\Delta z_{h} = 100$ \mbox{\AA}, tilting
        the field by $45^{\rm o}$ at $B_{\perp} = 4$ T increases
        the PL energy by $0.8$~meV. This exceeds the change in
         $\langle \omega \rangle$ due to the presence of Skyrmions
        in the strictly 2D system (Fig.~7). A measurement of the
        variation of $\langle \omega \rangle$ with $\theta$ may
         therefore be less suitable as a means of detecting
        Skyrmions in an asymmetric quantum well. However, because
         the subband energies are shifted {\em equally} on both
          sides of $\nu = 1$, the redshift of the LCP line remains
        unaffected by $B_{\parallel}$. The $\theta$ dependence of
        the redshift hence provides a means of detecting Skyrmions
         in an asymmetric quantum well.

         In a single heterojunction the spread of the hole wave
        function is {\em different} on either side of $\nu = 1$:
         for $\nu < 1$ the hole is essentially free (large $\Delta
        z_{h}$), while for $\nu > 1$ the hole is confined near the
         electron plane (smaller $\Delta z_{h}$) due to the
          formation of an interface exciton\citeup{Viet}. The
        diamagnetic shift of the hole subband energy ($HH_{0}$)
         for $\nu < 1$ then differs from the diamagnetic shift
          for $\nu > 1$. As a result, the redshift of the LCP line
         acquires an additional $\theta$ dependence given by
\begin{equation}
 \Delta \omega_{-} = \frac{\omega_{ch} (\Delta z_{h>})^{2} -
      \omega_{ch} (\Delta z_{h<})^{2} }{2 \ell_{\perp}^{2}}
\tan^{2} \theta ,
\end{equation}
         where $\Delta z_{h>}$ ($\Delta z_{h<}$) is the
          spread of the hole wave function for $\nu > 1$
         ($\nu < 1$). Since $\Delta z_{h>} < \Delta z_{h<}$
          the additional $\theta$ dependence leads to a
         decrease of the redshift with tilt angle, which
          may overwhelm the increase with tilt angle obtained
         for the strictly 2D system (see Fig.~8). This can
          explain the {\em decrease} of the redshift with
         tilt angle observed by Davies {\em et al.}\citeup{Davies}
        in their recent experiments on a single heterojunction.
        The additional $\theta$ dependence may be suppressed
         by applying a gate voltage to the heterojunction,
          which confines the hole equally on both sides of
          $\nu = 1$.

         The application of hydrostatic pressure also causes
         shifts in the electron and hole subband energies due
         to the different pressure dependences of the GaAs and
         Ga$_{1-x}$Al$_{x}$As bandgaps. This leads to an
         additional pressure dependence of the PL energy which
           does not occur in the strictly 2D system. For an
         asymmetric quantum well the redshift of the LCP line
          remains unaffected because the subband energies are
         shifted equally on both sides of $\nu = 1$.

\section{Optics with Skyrmions in Pure Systems}
       \noindent We now turn to a study of the optical properties
       of Skyrmions in pure systems. The initial state
       in a pure system is considerably different
        from that in a disordered system. In the absence of
      disorder the photoexcited hole is no longer localized
        near a minimum in the disorder potential. Thus
       Eq.~(\ref{eq:istate}) does not provide an adequate
        description of the initial state in a pure
       system.

       To get an idea of what the initial state {\em should}
         look like, consider the recombination of ordinary
       electrons and a holes in a pure system. The
        initial state in this case is the well-known
       magnetoexciton\citeup{Korolev}. The magnetoexciton
         has a quantum number ${\bf P}$, which plays the role
        of the total momentum in a magnetic field\citeup{Gorkov}.
       The energy spectrum consists of a series of bands
         $E_{nm}({\bf P})$, where $n$ and $m$ are the
       Landau-level indices of the electron and the
         hole, respectively\citeup{Lerner}. For an electron
        and a hole in the lowest Landau level the dispersion is
       $E_{00}({\bf P})$. Momentum conservation allows only
      magnetoexcitons with ${\bf P}$ equal to the photon wave
       vector to radiate. At optical frequencies the photon wave
      vector may be neglected. The PL spectrum is then determined
      by the recombination of magnetoexcitons with ${\bf P} = 0$.

       In light of the above considerations a number of
        questions arise with regard to the optical properties
       of Skyrmions in a pure system:
\begin{itemlist}
\item Can a Skyrmion bind with a hole to form a ``Skyrmionic''
        exciton?
\item Does such a Skyrmionic exciton have ${\bf P}$ as a good
       quantum number?
\item What is the dispersion relation of a Skyrmionic exciton?
\item What selection rules apply to the recombination of a
       Skyrmionic exciton?
\end{itemlist}
       Below we aim to answer these questions. We
       start by exposing the underlying symmetry that
      is responsible for the conservation of ${\bf P}$ in
        a pure system.

\subsection{Magnetic translation group}
       \noindent Conserved quantum numbers are a consequence of the
        invariance of a system under a group of symmetry operations.
       A well-known example is the angular momentum quantum
        number $m$, which results from invariance under rotations
       about the $z$ axis. The symmetry group associated with the
     quantum number ${\bf P}$ is the magnetic translation
      group\citeup{Zak}. Consider a
       particle of mass $m^{*}$ and charge $q$ in a magnetic field
      ${\bf B} = B \hat{z}$. Because the vector potential
      ${\bf A}({\bf r}) = \frac{1}{2} {\bf B} \times {\bf r}$
      depends on ${\bf r}$, the Hamiltonian $H = \frac{1}{2m^{*}}
        [-i \nabla - \frac{q}{c} {\bf A}({\bf r})]^{2}$
      is {\em not} invariant under ordinary translations
       $T_{\bf R} = e^{-{\bf R}\cdot \nabla}$. However, $H$ {\em is}
       invariant under {\em magnetic} translations
       $M_{\bf R} = e^{-i {\bf R} \cdot {\bf P}}$, where
      ${\bf P} = -i \nabla + \frac{q}{c} {\bf A}({\bf r})$.
        Using the Baker-Haussdorf theorem, the magnetic translation
       can be written as $M_{\bf R} = E_{\bf R} T_{\bf R}$, where
     $E_{\bf R} = \exp( i \frac{q}{c} {\bf A}({\bf R}) \cdot {\bf r})$
       is a gauge transformation. The gauge transformation compensates
       for the shift in the argument of the vector potential
        due to $T_{\bf R}$.

        In contrast to the ordinary translation group, the magnetic
        translation group is not Abelian. From the definition of
       $M_{\bf R}$ one can show that two successive magnetic
         translations satisfy the commutation relation
\begin{equation}
\label{eq:comrel}
M_{\bf R} M_{\bf R'} = e^{i \frac{q}{c} \Phi}
M_{\bf R'} M_{\bf R} ,
\end{equation}
       where $\Phi = {\bf B} \cdot ({\bf R} \times {\bf R}')$ is the
       flux enclosed by the parallellogram subtended by ${\bf R}$ and
     ${\bf R}'$. Equation~(\ref{eq:comrel}) lies at the heart of some
    remarkable phenomena exhibited by charged particles in a magnetic
      field, such as the Aharonov-Bohm effect\citeup{Aharonov},
      the Hofstadter butterfly\citeup{Hofstadter},
      and flux quantization in superconductors\citeup{Kittel}.
     Applying Eq.~(\ref{eq:comrel}) to infinitesimal magnetic
      translations yields the commutation relation for the
    generators of the magnetic translation group:
\begin{equation}
\label{eq:commut}
[P_{x},P_{y}] = -i \frac{q B}{c} .
\end{equation}
         Because the generators of the magnetic translation group
       do not commute, the energy eigenstates of a charged particle
       in a magnetic field can be chosen as eigenstates of {\em either}
       $P_{x}$ {\em or} $P_{y}$ (Landau gauge), or as eigenstates of
       $P_{x}^{2}+P^{2}_{y}$ (circular gauge), but not
        as eigenstates of {\em both} $P_{x}$ {\em and} $P_{y}$
       simultaneously. Hence ${\bf P}$ is not a good quantum number
        for a charged particle in a magnetic field.

      The commutator in Eq.~(\ref{eq:commut}) has the opposite sign for
       particles of opposite charge. The $x$ and $y$ components of
     ${\bf P} = -i \nabla_{1} - i \nabla_{2} + \frac{q}{c} {\bf A}
       ({\bf r}_{1}) - \frac{q}{c} {\bf A}({\bf r}_{2})$ therefore
      {\em do} commute with each other. Hence ${\bf P}$ {\em is} a
     good quantum number for two particles of opposite charge in a
     magnetic field. This is the classic result first obtained by
      Gor'kov and Dzyaloshinski\u{\i} for an exciton in a magnetic
     field\citeup{Gorkov}. Gor'kov and Dzyaloshinski\u{\i} further
     showed that ${\bf P}$ plays the role of the total momentum of
    the exciton
      in a magnetic field. This result can be understood in a simple
     way by considering the exciton not as two separate particles
       of opposite charge, but as one composite particle of zero
      charge. The motion of a neutral particle is unaffected by a
      magnetic field. Hence the center of mass of the exciton moves
    as a free particle whose energy eigenstates can be characterized
      by the total momentum ${\bf P}$. More generally it can be shown
     that for a system of $N$ charges with $\sum_{i=1}^{N} q_{i} = 0$,
      the $x$ and $y$ components of
\begin{equation}
\label{eq:gorkovp}
{\bf P} = \sum_{i=1}^{N} [-i \nabla + \frac{q_{i}}{c}
{\bf A}({\bf r}_{i})]
\end{equation}
      commute with each other. Thus the energy eigenstates of any
     neutral system of charges may be characterized by a conserved
      quantum number ${\bf P}$.

\subsection{Variational approach to Skyrmionic excitons}
      \noindent A Skyrmionic exciton consisting of a negatively
      charged Skyrmion and a hole {\em is} a charge neutral object.
     The Skyrmionic exciton must therefore have ${\bf P}$ as
      a good quantum number. Below we construct a variational
       wave function for a Skyrmionic exciton that has a
       definite value of ${\bf P}$.

     The initial state consisting of a Skyrmion and a hole
       localized at the origin [Eq.~(\ref{eq:istate})] can be
      written as $|i_{0}\rangle = a^{\dagger}_{0}|0\rangle$, where
\begin{equation}
\label{eq:istate_1}
a^{\dagger}_{0} = \prod_{m=0}^{\infty} (-u_{m}
e^{\dagger}_{m+1 \downarrow} e_{m \uparrow} + v_{m})
e^{\dagger}_{0 \downarrow} h^{\dagger}_{0 m_{j}} .
\end{equation}
      This state is not an eigenstate of ${\bf P}$. We construct an
     eigenstate of ${\bf P}$ by analogy with the tight-binding method
      for constructing Bloch states (which are characterized by a
      wave vector ${\bf k}$) from localized atomic
       orbitals\citeup{Ashcroft}. We first define a state
     $|i_{\bf R}\rangle = a^{\dagger}_{\bf R}|0\rangle$, where
\begin{equation}
\label{eq:istate_2}
a^{\dagger}_{\bf R} = M_{\bf R} a^{\dagger}_{0} M^{\dagger}_{\bf R} .
\end{equation}
      The state $|i_{\bf R}\rangle$ describes a Skyrmion and a
     hole localized at ${\bf R}$. In the absence of disorder
       the states $|i_{\bf R}\rangle$
     are degenerate with respect to ${\bf R}$. In the context of the
     tight-binding method this corresponds to the degeneracy of
      atomic orbitals on different sites. We form an extended
       state $|i_{\bf P}\rangle = a^{\dagger}_{\bf P}|0\rangle$
       by taking a linear superposition of localized states whose
      coefficients are plane waves:
\begin{equation}
\label{eq:istate_3}
a^{\dagger}_{\bf P} = \int d^{2} {\bf R} e^{i {\bf P} \cdot {\bf R}}
a^{\dagger}_{\bf R} .
\end{equation}
   Now consider the transformation properties of $|i_{\bf P}\rangle$
    under magnetic translation. Magnetic translation of an
    eigenstate of ${\bf P}$ gives back the same state multiplied
   by $e^{-i {\bf P} \cdot {\bf R}}$. Because magnetic translations
   act as an Abelian group on neutral excitations,
   $M_{\bf R} a^{\dagger}_{\bf R'} M_{\bf R} = a^{\dagger}_{\bf R+R'}$.
    Using this in Eq.~(\ref{eq:istate_3}) gives
\begin{equation}
M_{\bf R} a^{\dagger}_{\bf P} M^{\dagger}_{\bf R} =
e^{-i {\bf P} \cdot {\bf R}} a^{\dagger}_{\bf P} ,
\end{equation}
     which shows that $|i_{\bf P}\rangle$ is an eigenstate of ${\bf P}$.

      The state $|i_{\bf P}\rangle = a^{\dagger}_{\bf P}|0\rangle$
     given by Eqs.~(\ref{eq:istate_1})--(\ref{eq:istate_3}) describes
     a Skyrmionic exciton with generalized total momentum ${\bf P}$.
     It has an equal probability of finding the Skyrmion and the
      hole anywhere in the $x$-$y$ plane. Because $M_{\bf R}$ commutes
    with $H$ the states $|i_{\bf P}\rangle$ with different ${\bf P}$
     are orthogonal and uncoupled by $H$. The energy of the Skyrmionic
     exciton may be obtained by minimizing
\begin{equation}
\label{eq:sxenergy}
E_{\rm SX}({\bf P}) = \frac{\langle i_{\bf P}|H|i_{\bf P} \rangle}
                      {\langle i_{\bf P}|  i_{\bf P} \rangle} .
\end{equation}
     The values of $u_{m}$ and $v_{m}$ that minimize the energy will
    now depend on ${\bf P}$. For $u_{m} = 0$ and $v_{m} = 1$ we
    recover the magnetoexciton in the lowest Landau level\citeup{Lerner}.

\subsection{Energy of the Skyrmionic exciton}
     \noindent In this section we give an outline of the calculation
     of $E_{\rm SX}({\bf P})$ starting from
      Eq.~(\ref{eq:sxenergy}). The numerical results of the
      calculation are presented in Sec.~5.4 (for ${\bf P} = 0$)
    and Sec.~5.5 (for nonzero ${\bf P}$).

     We first calculate the denominator of Eq.~(\ref{eq:sxenergy}).
     Using Eq.~(\ref{eq:istate_3}), the denominator can be written as
\begin{equation}
\label{eq:norm}
\langle i_{\bf P}| i_{\bf P} \rangle =
S \int d^{2}{\bf R}\, e^{-i {\bf P} \cdot {\bf R}}
\langle i_{\bf R} | i_{0} \rangle ,
\end{equation}
      where $S$ is the sample area. In the coordinate
       representation $|i_{0} \rangle$ is the product of the hole
      state $\phi^{*}_{0}({\bf r})$ and a Slater determinant of electron
      orbitals $\psi_{m}({\bf r}) = -u_{m} \phi_{m+1}({\bf r})
      \chi_{\downarrow} + v_{m} \phi_{m}({\bf r}) \chi_{\uparrow}$,
     where $\chi_{\downarrow}$ and $\chi_{\uparrow}$ are spin wave
       functions. Similarly, $|i_{\bf R}\rangle$ is the product of the
      hole state $M^{\dagger}_{\bf R} \phi^{*}_{0}({\bf r})$ and a
     Slater determinant of electron orbitals $M_{\bf R} \psi_{m}({\bf r})$.
       The overlap between two Slater determinants is given by
      the determinant of the matrix of overlaps between the two
       sets of orbitals\citeup{Lowdin}. Hence
\begin{equation}
\label{eq:overlap}
\langle i_{\bf R}| i_{0} \rangle = e^{-\frac{|{\bf R}|^{2}}{4}}
 {\rm det}|A({\bf R})| ,
\end{equation}
      where $A({\bf R})$ is a matrix whose element
       $A_{mm'}({\bf R})$ is the overlap between
       $M_{\bf R} \psi_{m}({\bf r})$ and  $\psi_{m'}({\bf r})$.
      The factor in front of the determinant is the overlap
     between the hole states. Since $\chi_{\downarrow}$ and
      $\chi_{\uparrow}$ are orthogonal,
\begin{equation}
\label{eq:amatrix}
A_{m m'}({\bf R}) =
u^{*}_{m} u_{m'} \Delta_{m+1,m'+1}({\bf R}) +
v^{*}_{m} v_{m'} \Delta_{m m'} ({\bf R}) ,
\end{equation}
     where $\Delta_{m m'}({\bf R})$ is the overlap
      between $M_{\bf R} \phi_{m}({\bf R})$ and $\phi_{m'}({\bf r})$.
    The matrix $\Delta({\bf R})$ plays a crucial role in
    the calculation of the energy of the Skyrmionic exciton.
     Its matrix elements are given by (${\bf R} = X+i Y$)
\begin{equation}
\label{eq:dmatrix}
\Delta_{m m'}({\bf R}) =
\left\{ \begin{array}{ll}
\left(\frac{m!}{m'!}\right)^{1/2}
\left(\frac{{\bf R}^{*}}{\sqrt{2}}\right)^{m'-m}
e^{-\frac{|{\bf R}|^{2}}{4}} L^{m'-m}_{m}
\left(\frac{|{\bf R}|^{2}}{2}\right) & (m' \geq m) \\
\left(\frac{m'!}{m!}\right)^{1/2}
\left(-\frac{{\bf R}}{\sqrt{2}}\right)^{m-m'}
e^{-\frac{|{\bf R}|^{2}}{4}} L^{m-m'}_{m'}
\left(\frac{|{\bf R}|^{2}}{2}\right) & (m > m')
\end{array} \right. ,
\end{equation}
   where $L_{m}^{m'}$ is a Laguerre polynomial.

    In their current form, Eqs.~(\ref{eq:overlap})--(\ref{eq:dmatrix})
    do not provide a practical way of calculating
      $\langle i_{\bf R}|i_{0}\rangle$. This is because the size of
     the matrix $A({\bf R})$
    is determined by the degeneracy of the Landau level, which becomes
  infinite in the thermodynamic limit. However, on physical grounds we
   expect that $\langle i_{\bf R}|i_{0}\rangle$ should only depend on
     those $m$ states for which $u_{m}$ and $v_{m}$ are different
    from $0$ and $1$, respectively. This is indeed the case. Setting
   $u_{m} = 0$ and $v_{m} = 1$ for $m \geq M$, we can
     rewrite Eq.~(\ref{eq:overlap}) in terms of a matrix whose size is
     determined by $M$. The algebraic manipulations that are involved
      essentially amount to a particle-hole transformation, whereby the
     filled $\nu = 1$ state $|0\rangle$ becomes the vacuum state for
     spin-$\uparrow$ holes. The transformed expression {\em does}
    provide a practical way of calculating
       $\langle i_{\bf R}|i_{0}\rangle$.  A final Fourier transform
     then yields $\langle i_{\bf P}|i_{\bf P}\rangle$.

   Next we turn to the calculation of the numerator in
      Eq.~(\ref{eq:sxenergy}). Using Eq.~(\ref{eq:istate_3}),
    the numerator can be written as
\begin{equation}
\langle i_{\bf P} |H| i_{\bf P} \rangle =
A \int d^{2}{\bf R} e^{-i {\bf P} \cdot {\bf R}}
\langle i_{\bf R} |H| i_{0} \rangle .
\end{equation}
    To evaluate $\langle i_{\bf R}|H|i_{0}\rangle$ we use Wick's
     theorem\citeup{Mahan}. This theorem allows us to decompose
   the matrix element of a product of $n$ creation and
    $n$ annihilation operators between the states $|i_{0}\rangle$
     and $|i_{\bf R}\rangle$ into a product of ``propagators''
\begin{equation}
G_{m \sigma m' \sigma'}({\bf R}) =
\frac{ \langle i_{\bf R} | e^{\dagger}_{m \sigma}({\bf R})
e_{m' \sigma'} |i_{0} \rangle}{\langle i_{\bf R}| i_{0} \rangle} ,
\end{equation}
    where $e^{\dagger}_{m \sigma}({\bf R})$ creates an electron
     of spin $\sigma$ in the state $M_{\bf R} \phi_{m}({\bf r})$.
   The ``propagator'' $G_{m \sigma m' \sigma'}({\bf R})$ gives
    the probability amplitude that, if a hole is created in
   a state with angular momentum $m'$ and spin $\sigma'$
  centered on ${\bf R}' = 0$, it will ``propagate'' to a
   state with angular momentum $m$ and spin $\sigma$ centered
  on ${\bf R}' = {\bf R}$. In the coordinate representation
    the ``propagator'' is given by the overlap between two
   Slater determinants with one electron missing. This overlap
    can be expressed\citeup{Lowdin} in terms of the minors of
   the determinant of $A({\bf R})$. The minors in turn
    are related to the inverse of $A({\bf R})$. We thus obtain
\begin{eqnarray}
\label{eq:rdowndown}
G_{m \downarrow m' \downarrow}({\bf R}) & = &
u^{*}_{m-1} u^{*}_{m'-1} A^{-1}_{m'-1,m-1}({\bf R}) , \\
\label{eq:rupup}
G_{m \uparrow m' \uparrow}({\bf R}) & = &
v^{*}_{m} v_{m'} A^{-1}_{m' m}({\bf R}) , \\
\label{eq:rdownup}
G_{m \downarrow m' \uparrow}({\bf R}) & = &
u^{*}_{m-1} v_{m'} A^{-1}_{m-1, m'}({\bf R}) , \\
\label{eq:rupdown}
G_{m \uparrow m' \downarrow}({\bf R}) & = &
v^{*}_{m} u_{m'-1} A^{-1}_{m, m'-1}({\bf R}) .
\end{eqnarray}
   At ${\bf R} = 0$ we have $A^{-1}_{m m'}(0) = \delta_{mm'}$, and
    Eqs.~(\ref{eq:rdowndown})--(\ref{eq:rupdown})
   reduce to Eqs.~(\ref{eq:downdown})--(\ref{eq:downup}). The matrix
     element of $H$ is given by
\begin{equation}
\label{eq:sandwichr}
\frac{\langle i_{\bf R}|H|i_{0} \rangle}
     {\langle i_{\bf R}|  i_{0} \rangle} =
\varepsilon^{h}_{m_{j}} + U^{Z}({\bf R}) + U^{H}({\bf R}) -
         U^{ex}({\bf R}) - U^{sk}({\bf R}) - U^{eh}({\bf R}) ,
\end{equation}
    where $\varepsilon^{h}_{m_{j}}$ is the energy of the
    hole\fnm{d}\fnt{d}{Since in the absence of disorder
       $\varepsilon^{h}_{mm_{j}}$ is degenerate with respect to
     $m$ we have dropped this index.}, and
\begin{eqnarray}
\label{eq:uzr}
U^{Z}({\bf R}) & = & \frac{1}{2} |g_{e}| \mu_{B} B
 \sum_{m m'} \Delta_{m m'}({\bf R})(
G_{m \uparrow m' \uparrow}({\bf R}) -
G_{m \downarrow m' \downarrow}({\bf R}) + \delta_{m m'}) , \\
\label{eq:uhr}
U^{H}({\bf R}) & = & \frac{1}{2} \sum_{m m' m'' m'''}
V^{ee}_{m m' m'' m'''}({\bf R})
(G_{m \downarrow m''' \downarrow}({\bf R}) +
G_{m \uparrow m''' \uparrow}({\bf R}) - \delta_{m m'''})
\nonumber \\ & & \mbox{} \times (
G_{m' \downarrow m'' \downarrow}({\bf R}) +
G_{m' \uparrow m'' \uparrow}({\bf R}) - \delta_{m' m''}) , \\
\label{eq:uexr}
U^{ex}({\bf R}) & = & \frac{1}{2} \sum_{m m' m'' m'''}
V^{ee}_{m m' m'' m'''}({\bf R})(
G_{m \downarrow m'' \downarrow}({\bf R})
G_{m' \downarrow m'' \downarrow}({\bf R})
\nonumber \\ & & \mbox{} +
G_{m \uparrow m'' \uparrow}({\bf R})
G_{m'\uparrow m'''\uparrow}({\bf R}) - \delta_{m m''}
                                       \delta_{m' m'''} ) , \\
\label{eq:uskr}
U^{sk}({\bf R}) & = & \sum_{m m' m'' m'''}
V^{ee}_{m m' m''-1\,m'''+1}({\bf R})
G_{m \downarrow m''-1\,\uparrow}({\bf R})
G_{m' \uparrow m'''+1\,\downarrow}({\bf R}) , \\
\label{eq:uehr}
U^{eh}({\bf R}) & = & \sum_{m m'} V^{eh}_{m 0 0 m'}({\bf R})(
 G_{m \downarrow m' \downarrow}({\bf R}) +
 G_{m \uparrow m' \uparrow}({\bf R}) - \delta_{m m'}) ,
\end{eqnarray}
    with
\begin{eqnarray}
\label{eq:veer}
V^{ee}_{m m' m'' m'''}({\bf R}) & = & \sum_{n n'}
V^{ee}_{n n' m'' m'''}\Delta_{mn}({\bf R}) \Delta_{m'n'}({\bf R}), \\
\label{eq:vehr}
V^{eh}_{m m' m'' m'''}({\bf R}) & = & \sum_{n n'}
V^{eh}_{n n' m'' m'''}\Delta_{mn}({\bf R}) \Delta^{*}_{m' n'}({\bf R}) .
\end{eqnarray}
     One may check that at ${\bf R} = 0$ each of the terms in
      Eqs.~(\ref{eq:uzr})--(\ref{eq:uehr}) reduces to the
     corresponding term in Eqs.~(\ref{eq:uz})--(\ref{eq:ueh}).
       After a suitable particle-hole transformation [see the
      discussion after Eq.~(\ref{eq:dmatrix})],
       Eqs.~(\ref{eq:sandwichr})--(\ref{eq:vehr}) provide a
       practical (though cumbersome) way of calculating
      $\langle i_{\bf R}|H|i_{0}\rangle$. Taking the Fourier
      transform yields $\langle i_{\bf P}|H|i_{\bf P} \rangle$.
     The calculation is completed by numerically minimizing
     $E_{\rm SX}({\bf P})$ as a function of the parameters
      $u_{m}$ and $v_{m}$ with $m < M$.

\subsection{Zero-momentum state}
     \noindent The initial state with ${\bf P} = 0$ is of
     particular interest because it is the state that determines
      the recombination spectrum in a disorderless system.
        Figure~9(a) shows the
     difference between the energy of a Skyrmionic exciton
      with ${\bf P} = 0$ and the energy of a magnetoexciton
     with ${\bf P} = 0$. Figure~9(b) shows the corresponding
       difference in spin.
      Figure~9 should be compared with the equivalent
       Fig.~6 for a Skyrmion bound to a localized hole.
       The two figures are very similar in their physical trends.
      The size of the Skyrmion bound to the
       itinerant hole is determined by the same three competing
      effects that determine the size of the Skyrmion bound
      to the localized hole (Sec.~4.3). Again the Skyrmion
       shrinks with increasing Zeeman energy, and when the
      hole is brought closer to the electron plane. The threshold
     value of $g$ below which Skyrmions form at a given $d$ is
     almost the same in both figures.

       The energy and spin differences shown in Fig.~9 were
       obtained using a variational wave function with $M = 14$
     parameters. This accounts for the discrepancy between the
     energy scales in Fig.~9 and Fig.~6 (which was obtained using
    a wave function with $M = 60$ parameters). In Fig.~10 we compare
     the energy and spin differences for the delocalized state
     $|i_{{\bf P} = 0}\rangle$ with the energy and spin differences
    for the localized state $|i_{{\bf R} = 0}\rangle$, using the
    {\em same} number of parameters ($M = 14$) for both states.
       Ideally we would like to use as many parameters ($M = 60$)
       for the delocalized state as for the localized state.
        Unfortunately the increased complexity of the calculation
      limits us at present to a maximum of $M = 14$ parameters.
       We believe that this number of parameters is sufficient
        to capture the essential features. We have found that
      increasing the number of parameters from $M = 10$ to $M = 14$
       increases the absolute value of $\Delta E$ and $\Delta S_{z}$,
      but does not alter their behaviour with $g$ and $d$. We
      therefore expect that the results of a calculation
       with $M = 60$ parameters will be similar to those
      shown in Fig.~9, except for a change of scale along
      the vertical axes. Figure~10 demonstrates that for wave
        functions with the same number of parameters, the
      delocalized state has a lower energy and a larger spin than
       the localized state. From our results for the localized state
      using $M = 60$ parameters (Fig.~6) we therefore expect
       Skyrmionic excitons with at least $3$--$4$ spin flips in
       systems of experimental interest.

       Most importantly, our calculation shows that the
        lowest-energy state probed by a PL experiment at $\nu = 1$
       is {\em not} a ${\bf P} = 0$ magnetoexciton. Instead, the
      lowest-energy state is a ${\bf P} = 0$ Skyrmionic exciton.
       In the following section we explore the consequences of
       this result for the optical recombination spectrum.

\subsection{Optical recombination of Skyrmionic excitons}
      \noindent The selection rules for the optical recombination
      in a pure system require the conservation of the
       total generalized momentum ${\bf P}$. The conservation
     of ${\bf P}$ arises from the invariance of the pure
      system under magnetic translation, and is not required
      in the disordered system. In addition we still have
       the spin selection rule, which stipulates that the
      final state must contain $|S_{z} + \frac{1}{2}|$ reversed
     spins in the case of RCP recombination, and
       $|S_{z} - \frac{1}{2}|$ reversed spins in the case of
      LCP recombination. The reversed spins in the pure
      system are not simple spin flips, but spin waves which
        have their own momentum quantum number ${\bf P}$.

       The selection rules allow us to make an important observation
      about the recombination spectrum without having to calculate
       the detailed line shape. Assuming (as usual) complete energy
         relaxation before recombination, the initial state is a
       Skyrmionic exciton with ${\bf P} = 0$ and spin $S_{z}$. The
        selection rules then require a final state containing
         $|S_{z} \pm \frac{1}{2}|$ spin waves, whose {\em total}
        generalized momentum is ${\bf P} = 0$. For final states
        containing multiple spin waves, the final state energies
       form a continuum between $(E_{f})_{min} = |S_{z} \pm \frac{1}{2}|
        |g_{e}| \mu_{B}B$ (corresponding to $|S_{z} \pm \frac{1}{2}|$
       spin waves, {\em each} of zero momentum), and $(E_{f})_{max} =
       |S_{z} \pm \frac{1}{2}| (|g_{e}|\mu_{B}B + \sqrt{\pi/2}\,
        e^{2}/\epsilon \ell)$ (corresponding to $|S_{z} \pm \frac{1}{2}|$
        widely separated spin-$\downarrow$ quasielectrons and
      spin-$\uparrow$ quasiholes). Here we have used Larmor's theorem,
       which says that the energy of a spin wave of zero momentum is
        given by $|g_{e}|\mu_{B}B$, regardless of the interaction among
       the electrons. Hence the RCP and LCP lines have an {\em intrinsic}
       width $|S_{z} \pm \frac{1}{2}| \sqrt{\pi/2}\,e^{2}/\epsilon \ell$,
        where the upper sign applies to RCP and the lower sign to
        LCP\@. The LCP line has a larger intrinsic width than the
        RCP line since more spin waves are left in the final state.

       The intrinsic width is a unique feature of the recombination
        spectrum of a Skyrmionic exciton. A similar broadening of the
       line shape due to the formation of a spin texture has been
        found in the absorption spectrum of a Skyrmion bound to a
         charged impurity\citeup{Cote}. The recombination spectrum
       of a magnetoexciton has no intrinsic width because the final
       state is either the fully polarized state $|0\rangle$ (in the
         case of RCP), or a state containing one spin wave of zero
        momentum (in the case of LCP). In either case there is only
         one final-state energy, and hence the spectrum consists of
        a sharp line in both polarizations.

       An additional {\em extrinsic} broadening of the luminescence
       lines can occur due to finite temperatures or the presence of
     a weak disorder. The extrinsic
       broadening of the recombination spectrum of a magnetoexciton has
      been examined in detail by Cooper and Chklovskii\citeup{Cooper}.
       They have shown that the extrinsic broadening can account
        for the difference in line width of the RCP and LCP lines
       observed by Plentz {\em et al.}\citeup{Plentz} in narrow
       quantum wells. The observed difference in line width is also
        consistent with the intrinsic broadening characteristic of
        Skyrmionic excitons, provided they could form in these narrow
       wells. In wide quantum wells both the intrinsic and the
        extrinsic mechanisms will contribute to the broadening of
       the luminescence lines. It would be interesting to study the
        evolution of the line width with
        temperature and disorder. The difference in line width should
           persist to low temperatures and high mobilities in wide
         wells, where the intrinsic mechanism dominates, but not
          in narrow wells, where the extrinsic mechanism operates.

       The PL energies for the pure system
       can be calculated using the method of moments discussed
       in Sec.~4.4. Instead of the localized state $|i_{0}\rangle$,
       we now use the delocalized state $|i_{\bf P}\rangle$ in
       Eq.~(\ref{eq:moments}). We have calculated the energies
       of the RCP and LCP lines for transitions from the
      ${\bf P} = 0$ state. The results are qualitatively similar
       to the results for the disordered system shown in
      Fig.~7. The mean luminescence energy
       $\langle \omega \rangle$ increases with $g$ up to a
       threshold $g_{c}$. Above the threshold $\langle \omega \rangle$
      does not vary with $g$. By setting $u_{m} = 0$ and $v_{m} = 1$
       we can show that for a magnetoexciton $\langle \omega \rangle$
      is independent of $g$. Hence a measurement of the $g$ dependence
      of $\langle \omega \rangle$ could be used to distinguish the
       optical recombination of a Skyrmionic exciton from the
       optical recombination of a magnetoexciton. However, due
        to the finite-size effect (Sec.~4.7) the $g$ dependence of
       the redshift of the LCP line might provide a clearer
        optical signature. For a Skyrmionic exciton the redshift
       increases with $g$ while for a magnetoexciton the redshift
       is independent of $g$. Using a wave function with $M = 14$
        parameters, we find that the redshift increases by $0.03$
         $e^{2}/\epsilon \ell$ when $g$ varies between $0.025$
        and $0.02$ (here $d = 3 \ell$). The change in the redshift
         may be measured by the methods discussed in Sec.~4.6.

\subsection{Dispersion relation}
     \noindent Thus far our discussion has focused on the state with
       ${\bf P} = 0$, which determines the PL spectrum of the
      pure system at low temperatures.
     States with nonzero ${\bf P}$ can also be probed, for example
      by means of a resonant light scattering experiment. In this
     section we derive exact results for the electric dipole moment
     and the asymptotic form of the dispersion of a Skyrmionic
      exciton. We also discuss features of the dispersion relation
       obtained from the variational state $|i_{\bf P}\rangle$.

       The electric dipole moment of a Skyrmionic exciton is
        given by
\begin{equation}
\label{eq:dipole}
     {\bf d} = \frac{c}{B^{2}} {\bf P} \times {\bf B} .
\end{equation}
       The electric dipole moment is independent of the
       spin, and equal to the dipole moment of
      a magnetoexciton in the lowest Landau level\citeup{Lerner}.
       Equation~(\ref{eq:dipole}) is a general result for a
        Skyrmionic exciton restricted to the lowest Landau level,
       which does not rely on the specific form of the variational
         wave function in
       Eqs.~(\ref{eq:istate_1})--(\ref{eq:istate_3}). To prove
        Eq.~(\ref{eq:dipole}), we write the Gor'kov momentum
       for $N$ charges in a magnetic field as
\begin{equation}
\label{eq:pgorkov}
{\bf P} = \sum_{i=1}^{N} {\bf \Pi}_{i} + \frac{1}{c}
{\bf B} \times {\bf d} ,
\end{equation}
        where ${\bf \Pi}_{i} = -i \nabla_{i} - \frac{q_{i}}{c}
      {\bf A}({\bf r}_{i})$ is the dynamical momentum of
        charge~$i$, and ${\bf d} = \sum_{i=1}^{N} q_{i}
       {\bf r}_{i}$ is the electric dipole moment. The quantities
          ${\bf P}$, ${\bf \Pi}_{i}$, and
        ${\bf d}$ should at
        this stage be regarded as operators. We
       now take the expectation value of Eq.~(\ref{eq:pgorkov})
       in an eigenstate of the operator ${\bf P}$. For a state
      in the lowest Landau level the expectation value of
       $\sum_{i=1}^{N} {\bf \Pi}_{i}$ is equal to zero. Hence
        ${\bf P} = \frac{1}{c} {\bf B} \times {\bf d}$, where
       ${\bf P}$ and ${\bf d}$ are now the expectation values
       of the corresponding operators. Taking the cross product
         with ${\bf B}$ proves Eq.~(\ref{eq:dipole}).

       We can use Eq.~(\ref{eq:dipole}) to establish the exact
        asymptotic behaviour of the dispersion relation for a
     Skyrmionic exciton in the limit $|{\bf P}| \rightarrow \infty$.
        Suppose that, instead of Eq.~(\ref{eq:psi+}), we were
       given the {\em exact} ground state for a Skyrmion localized
        at the origin of the $x$-$y$ plane. Such an exact ground
       state has been found explicitly for the case of a hard-core
      interaction among the electrons\citeup{MacDonald}.
          We denote the exact
        ground state by $s_{0}^{\dagger}|0\rangle$, and its
       energy by $\varepsilon_{\rm SK}$.  We now use the state
        $a^{\dagger}_{0}|0\rangle = s^{\dagger}_{0}
         h^{\dagger}_{0 m_{j}}|0\rangle$ to construct the
        Skyrmionic exciton, instead of Eq.~(\ref{eq:istate_1}).
       The resulting state $|i_{\bf P}\rangle$ is an exact
         eigenstate of $H^{e} + H^{h} + V^{ee}$ with eigenvalue
         $\varepsilon^{h}_{m_{j}} + \varepsilon_{\rm SK}$. If
        $s^{\dagger}_{0}|0\rangle$ is within the lowest Landau
         level, $|i_{\bf P}\rangle$ has an electric dipole moment
        given by Eq.~(\ref{eq:dipole}). The electric dipole moment
          may have two contributions: one from the separation
            between the hole and the center of mass of the Skyrmion,
         and one from the internal dipole moment of the Skyrmion
          relative to its center of mass. However, because the
        Skyrmion is in its ground state the internal dipole moment
          must vanish.
            Hence the separation between the hole and center
           of mass of the Skyrmion in the state $|i_{\bf P}\rangle$
         is ${\bf r}_{0} = {\bf d}/e$. The electron-hole interaction
          $V^{eh}$ breaks the degeneracy of the states
           $|i_{\bf p}\rangle$ with respect to ${\bf P}$. For large
          $|{\bf P}|$ the energy shift due to $V^{eh}$ is
          $-e^{2}/(\epsilon |{\bf r}_{0}|)$. Thus the exact
           asymptotic behaviour of the dispersion in the limit
            $|{\bf P}| \rightarrow \infty$ is
\begin{equation}
\label{eq:limit}
E_{\rm SX}({\bf P}) \sim
\varepsilon^{h}_{m_{j}} +  \varepsilon_{\rm SK} -
\frac{e^{2}}{\epsilon |{\bf P}| \ell^{2}}
\end{equation}

           An approximate dispersion law for the Skyrmionic exciton
          is obtained by minimizing the energy of the variational
        state given by Eqs.~(\ref{eq:istate_1})--(\ref{eq:istate_3})
         as a function of ${\bf P}$. The results for a variational
          state with $M = 10$ parameters are shown in Fig.~11.
           Near ${\bf P} = 0$ the dispersion is parabolic with an
          effective mass that differs by less than 7~\% from the
          effective mass of a magnetoexciton. As $|{\bf P}|$
            increases, the energy of the Skyrmionic exciton
           approaches that of a magnetoexciton. Above a certain
            value of $|{\bf P}|$ (approximately $3.5/\ell$ for
      the values of $g$ and $d$ chosen in Fig.~11) the magnetoexciton
          is the lowest-energy state within the variational space.

         While the variational state gives sensible results
          for the dispersion at small values of $|{\bf P}|$, this
          state does not adequately describe the dispersion at
          large $|{\bf P}|$. The dispersion obtained from the
        variational state approaches $\varepsilon^{h}_{m_{j}} +
         \frac{1}{2}|g_{e}|\mu_{B}B$ as $|{\bf P}| \rightarrow \infty$,
        whereas we know from Eq.~(\ref{eq:limit}) that the exact
         dispersion must approach $\varepsilon^{h}_{m_{j}} +
           \varepsilon_{\rm SK}$. The failure of the variational state
          occurs because the Skyrmion develops a large internal
         dipole moment at high $|{\bf P}|$.
       To illustrate this point we calculate the pair correlation
         function $g_{\downarrow \uparrow}({\bf r})$ for the simplest
       variational state with $M = 1$. This state allows
          only one extra spin flip. The spin flip
           creates an additional spin-$\downarrow$ electron and
         a spin-$\uparrow$ hole. The pair correlation function
          $g_{\downarrow \uparrow}({\bf r})$ is proportional to
           the probability of finding a spin-$\downarrow$ electron
           and a spin-$\uparrow$ hole separated by ${\bf r}$.
         The result is
\begin{equation}
\label{eq:gexc}
2 \pi g_{\downarrow \uparrow}({\bf r}) = \frac{u^{2}_{0}}
{u^{2}_{0} + \frac{1}{2} v^{2}_{0} e^{-\frac{1}{4} {\bf P}^{2}}}
 \left(\frac{4}{9} + \frac{2}{27} ({\bf r} - \frac{1}{2} \hat{z} \times
{\bf P})^{2} \right) e^{ -\frac{1}{3}({\bf r} - \frac{1}{2} \hat{z}
\times {\bf P})^{2}} .
\end{equation}
        The average separation between the
      spin-$\downarrow$ electrons and the spin-$\uparrow$
       hole is
\begin{equation}
\label{eq:rexc}
{\bf r}_{\downarrow \uparrow} = \frac{u^{2}_{0}}
{2 u^{2}_{0} + v^{2}_{0} e^{-\frac{1}{4}
{\bf P}^{2}}}\: \hat{z} \times {\bf P} .
\end{equation}
       For comparison, the pair correlation function for the
         localized Skyrmion is
\begin{equation}
\label{eq:gloc}
2 \pi g_{\downarrow \uparrow}({\bf r}) =
u^{2}_{0}\left( \frac{3}{4} + \frac{1}{16} {\bf r}^{2} \right)
e^{-\frac{1}{4} {\bf r}^{2}} ,
\end{equation}
      and ${\bf r}_{\downarrow \uparrow} = 0$.
       Equation~(\ref{eq:rexc}) shows that the average separation
      between the spin-$\downarrow$ electrons and the
         spin-$\uparrow$ hole becomes larger as $|{\bf P}|$ is
        increased. Thus the Skyrmion develops an internal
       dipole moment.

       The internal dipole moment which is associated
      with flipping an extra spin in the state $|i_{\bf P}\rangle$
       becomes energetically unfavorable at high $|{\bf P}|$. Hence
       the magnetoexciton becomes
      the lowest-energy state above a certain value of $|{\bf P}|$.
       According to the arguments given earlier, the internal
     dipole moment must vanish in the exact state $|i_{\bf P}\rangle$.
       The variational state thus fails to reproduce the correct
      asymptotic behaviour of the dispersion at large $|{\bf P}|$.
       Increasing the number of variational parameters cannot
        suppress the internal dipole moment and thus does not
       remedy this failure. The construction of a
        variational state that {\em does} produce the correct
       dispersion in the high-$|{\bf P}|$ limit is left for
        for future research.

\section{Summary}
       \noindent In this review we have presented a detailed
         investigation of the optical properties of quantum Hall
         Skyrmions. Skyrmions replace ordinary spin-$\frac{1}{2}$
          quasiparticles
        as the elementary charged excitations of a quantum Hall
         system at filling factor $\nu = 1$ for sufficiently
           small Zeeman coupling. The optical recombination
        of Skyrmions with photoexcited holes determines
       the PL spectrum near $\nu = 1$ in heterojunctions
         and quantum wells where the electron-hole separation exceeds
        a critical value.

         The optical recombination in a disordered system takes
       place between a Skyr\-mion and a hole localized near a minimum
         in the disorder potential. The PL spectrum
        at $\nu \geq 1$ consists of a RCP line and a LCP line whose
       mean energy: (1) does not depend on $g$ for spin-$\frac{1}{2}$
        quasielectrons, (2) does depend on $g$ for Skyrmions. At
          $\nu < 1$ the spectrum consists of a LCP line shifted
         down in energy from the LCP line at $\nu \geq 1$. The $g$
        dependence of the redshift provides a way to determine
        the nature of the negatively charged excitations using
         PL spectroscopy. The redshift increases
        with $g$ if the charged excitations are Skyrmions, but
        does not depend on $g$ is they are spin-$\frac{1}{2}$
         quasielectrons. The $g$ dependence of the redshift can
         be measured, for example, by tilting the magnetic field
        or by applying hydrostatic pressure.

          Skyrmionic excitons---consisting of a Skyrmion and a
           hole bound together by their mutual Coulomb
         attraction---govern the optical recombination in a
          pure system. Skyrmionic excitons replace
         ordinary magnetoexcitons as the lowest-energy excitations
          probed by a PL experiment at $\nu = 1$.
         The $g$ dependence of the redshift of the LCP line
        can be used to distinguish between the optical recombination
         of a Skyrmionic exciton and the optical recombination
         of a magnetoexciton. Skyrmionic excitons have their own
         dispersion relation in terms of a quantum number
         ${\bf P}$, which plays the role of the total momentum in
        a magnetic field. The electric dipole moment of a Skyrmionic
         exciton is independent of its spin and equal to the dipole
         moment of a magnetoexciton.

        It is hoped that the results presented in this review
         will aid in the interpretation of PL
        experiments on quantum Hall systems near filling factor
         $\nu = 1$. They might also be of relevance to spectroscopic
        studies at other filling factors for which the ground state
         is spin polarized. For a more detailed comparison between
        theory and experiment a calculation of the line shape is
        needed. Efforts in this direction are currently underway.

\nonumsection{Acknowledgements}
  \noindent We thank Andrew Turberfield, Darren Leonard,
      Georg Bruun, Igor Lerner, and Stuart Trugman for helpful
      discussions. This work was supported by EPSRC Grant
      No.\ GR/K 15619.

\begin{figure}[p]
\mbox{\tenbf Figures}
\end{figure}
\begin{figure}[p]
\fcaption{Spin profile of (a) a negatively charged Skyrmion,
(b) a positively charged Skyrmion. (Adapted from Ref.~19.)}
\end{figure}
\vspace*{-1em}
\begin{figure}[p]
\fcaption{(a) Energy, and (b) spin of a negatively
charged Skyrmion as a function of $g = \frac{1}{2}|g_{e}|\mu_{B}B/
(e^{2}/\epsilon \ell)$. Dash-dotted line: energy of a spin-$\frac{1}{2}$
quasielectron.}
\end{figure}
\vspace*{-1em}
\begin{figure}[p]
\fcaption{Band diagram of a one-side modulation-doped quantum well.
   The lowest electron ($E_{0}$) and hole ($HH_{0}$) energy levels
   and $z$ wave functions are shown. Dotted lines indicate the
    average positions of electrons and holes along $z$.}
\end{figure}
\vspace*{-1em}
\begin{figure}[p]
\fcaption{
   Magnetic-field dependence of the $E_{0}$ luminescence energy
     from a GaAs/Ga$_{0.68}$Al$_{0.32}$As heterojunction with
    electron density $n_{s} = 9.7$ $\times$ $10^{10}$ ${\rm cm}^{-2}$
    and mobility $\mu = 9$ $\times$ $10^{6}$ ${\rm cm}^{2}
     {\rm V}^{-1} {\rm s}^{-1}$ at $T = 120$ mK\@. Inset shows
    the luminescence spectra at selected magnetic fields.
    (Reprinted with kind permission from Andrew Turberfield.)}
\end{figure}
\vspace*{-1em}
\begin{figure}[p]
\fcaption{Comparison of PL at $\nu = 1$
(left column) and $\nu = 1^{-}$ (right column).
(a) Ground state before photoexcitation.
(b) Initial state before recombination.
(c) Final state after LCP recombination.
(d) Final state after RCP recombination.}
\end{figure}
\vspace*{-1em}
\begin{figure}[p]
\fcaption{
Difference in (a) energy and (b) spin between an initial state
consisting of a Skyrmion and a localized hole, and an initial
state consisting of a spin-$\downarrow$ electron and a localized
hole against a $\nu = 1$ background.}
\end{figure}
\vspace*{-1em}
\begin{figure}[p]
\fcaption{
    Mean luminescence energy $\langle \omega \rangle$ as a function
   of $g$, for several values of the separation $d$ between the
    electron and hole planes. Remaining parameter values are for
   a GaAs/Ga$_{1-x}$Al$_{x}$As heterojunction with electron
   density $n_{s} = 10^{11}$ ${\rm cm}^{-2}$. The gap value
   $E_{g} = 1509$ meV was taken from Ref.~21.}
\end{figure}
\vspace*{-1em}
\begin{figure}[p]
\fcaption{Variation of the redshift of the LCP line with
tilting angle ($\theta$) and applied hydrostatic pressure ($p$)
for a GaAs/Ga$_{1-x}$Al$_{x}$As quantum well or heterojunction
with $n_{s} = 10^{11}$ cm$^{-2}$ and a separation of
$d = 400$ \mbox{\AA} between the electron and hole planes.}
\end{figure}
\vspace*{-1em}
\begin{figure}[p]
\fcaption{
Difference in (a) energy and (b) spin between a
Skyrmionic exciton with ${\bf P} = 0$ and a
magnetoexciton with ${\bf P} = 0$. The energy and spin
are plotted as a function of the reduced Zeeman energy $g$,
for several values of the separation $d$ between the electron
and hole planes. The number of variational parameters is
$M = 14$.}
\end{figure}
\vspace*{-1em}
\begin{figure}[p]
\fcaption{
Comparison of (a) energy and (b) spin differences for the
delocalized state $|i_{{\bf P} = 0}\rangle$ and the
localized state $|i_{{\bf R} = 0}\rangle$. The energy and
spin differences were calculated using the same number of
variational parameters ($M = 14$) for both the delocalized
and the localized state. The separation between the electron
and hole planes is $d = 3\ell$.}
\end{figure}
\vspace*{-1em}
\begin{figure}[p]
\fcaption{Solid line: dispersion relation for the Skyrmionic exciton
(SX) obtained from a variational state with $M = 10$ parameters.
Dash-dotted line: dispersion relation for a magnetoexciton (MX)
in the lowest Landau level. $g = 0.005$ and $d = 3\ell$.}
\end{figure}

\clearpage
\nonumsection{References}

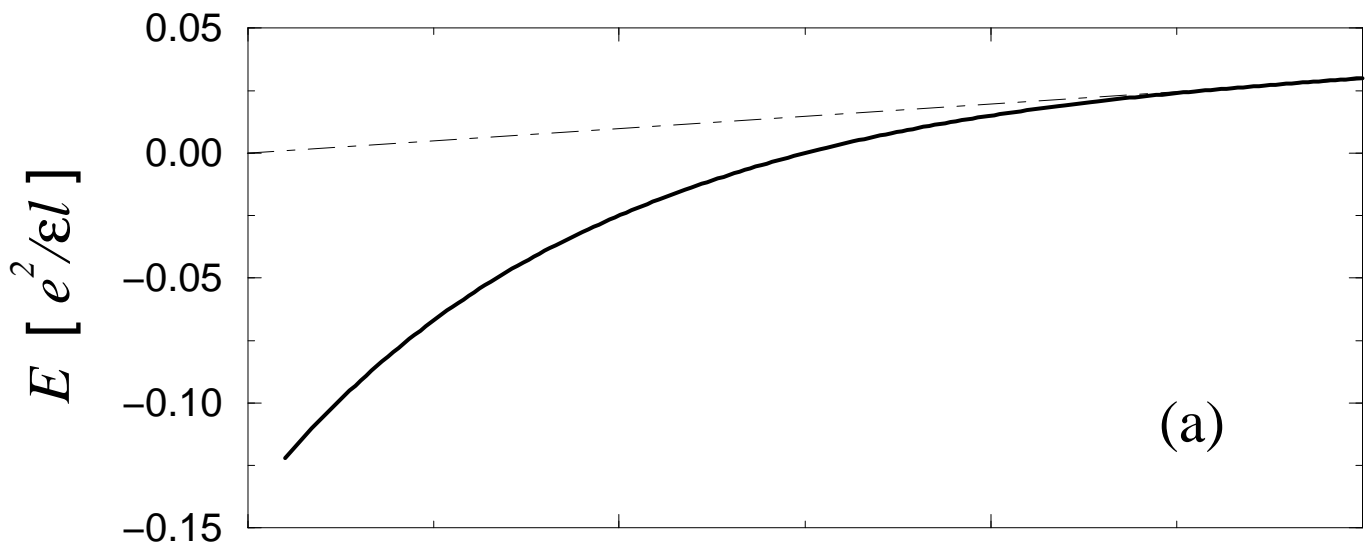
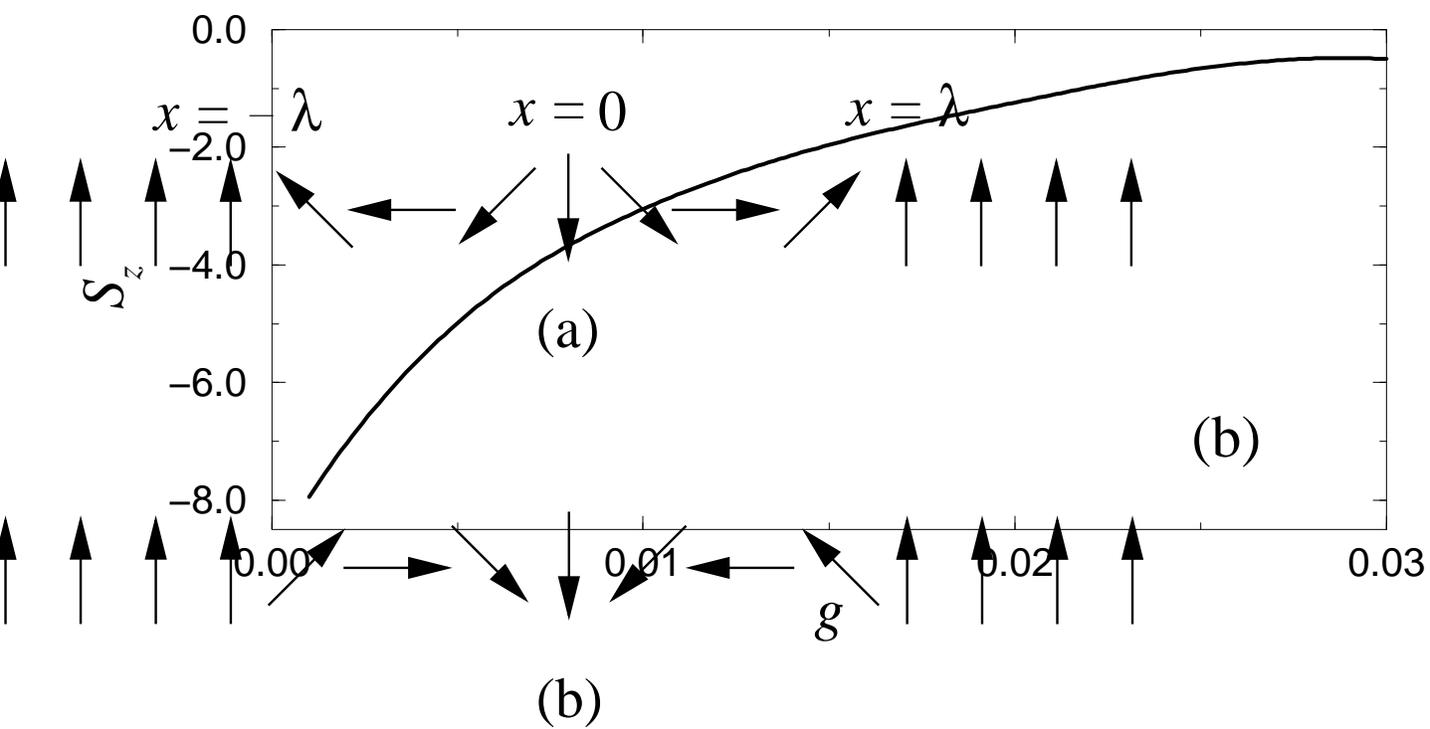
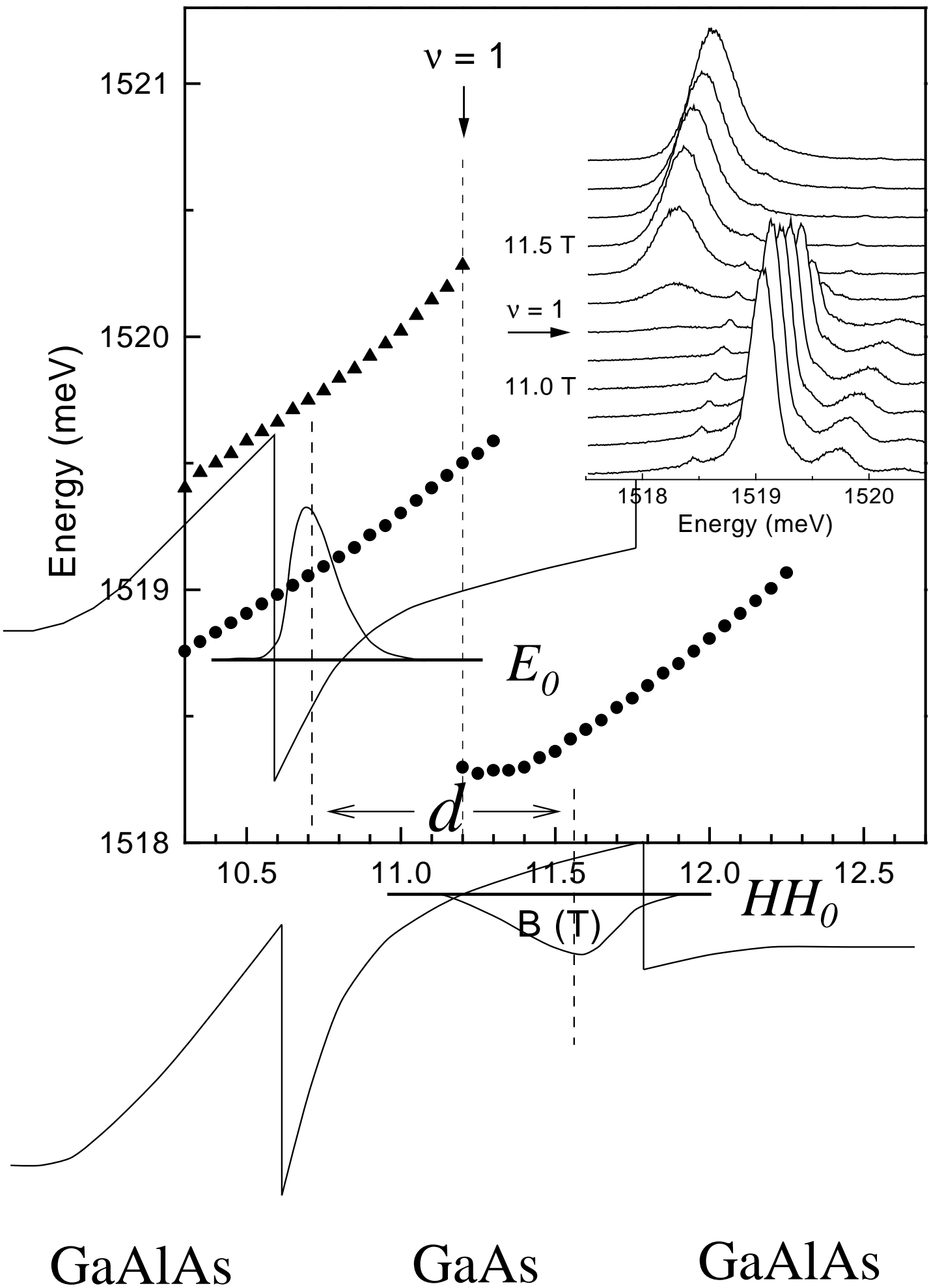
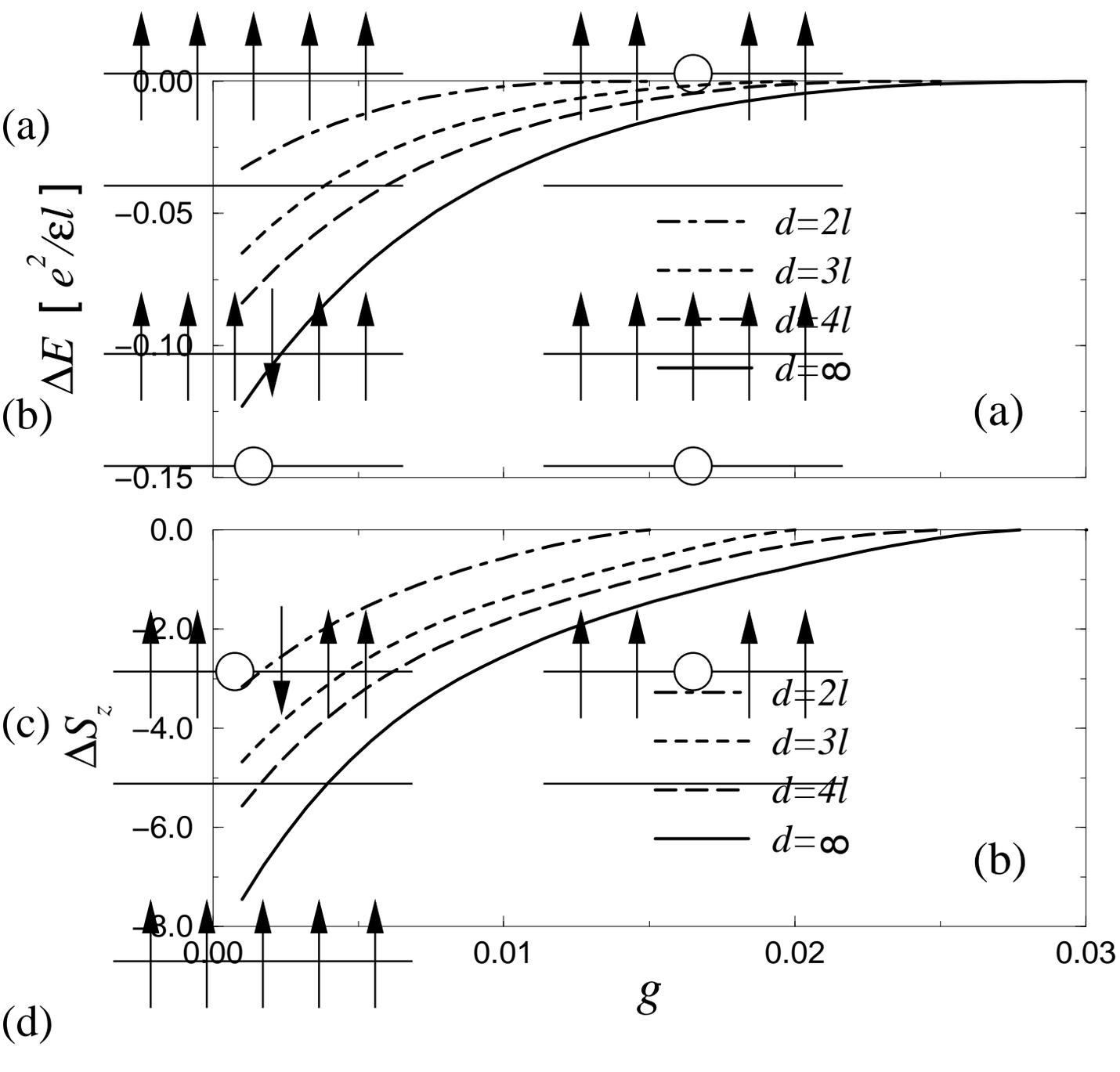
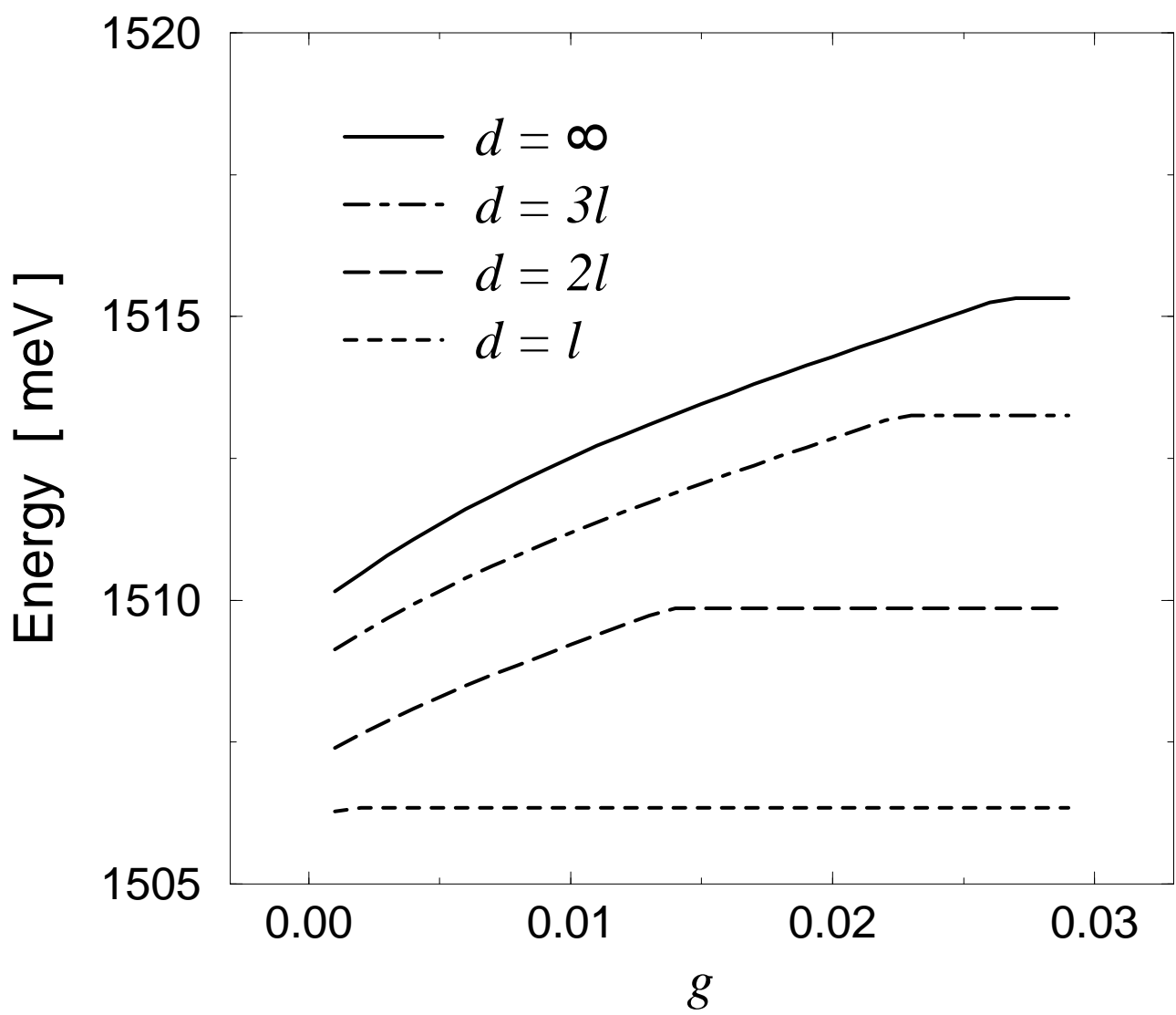
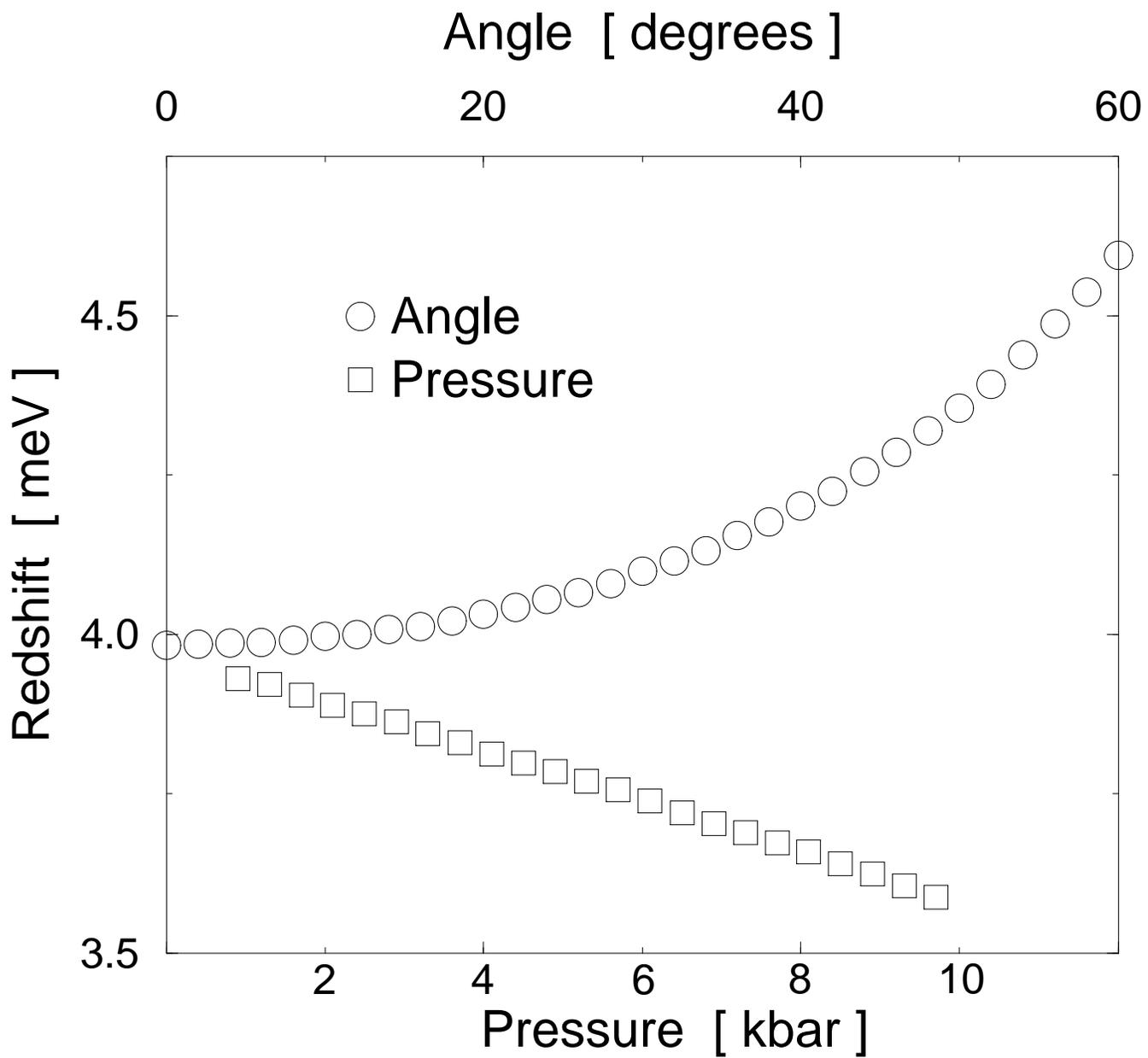
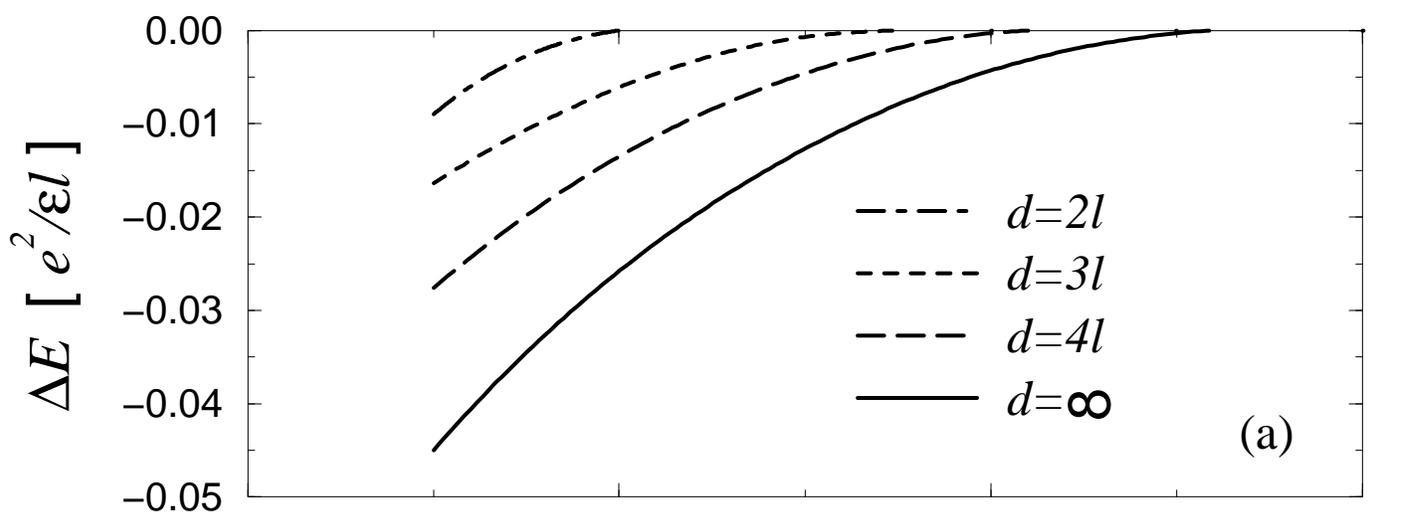
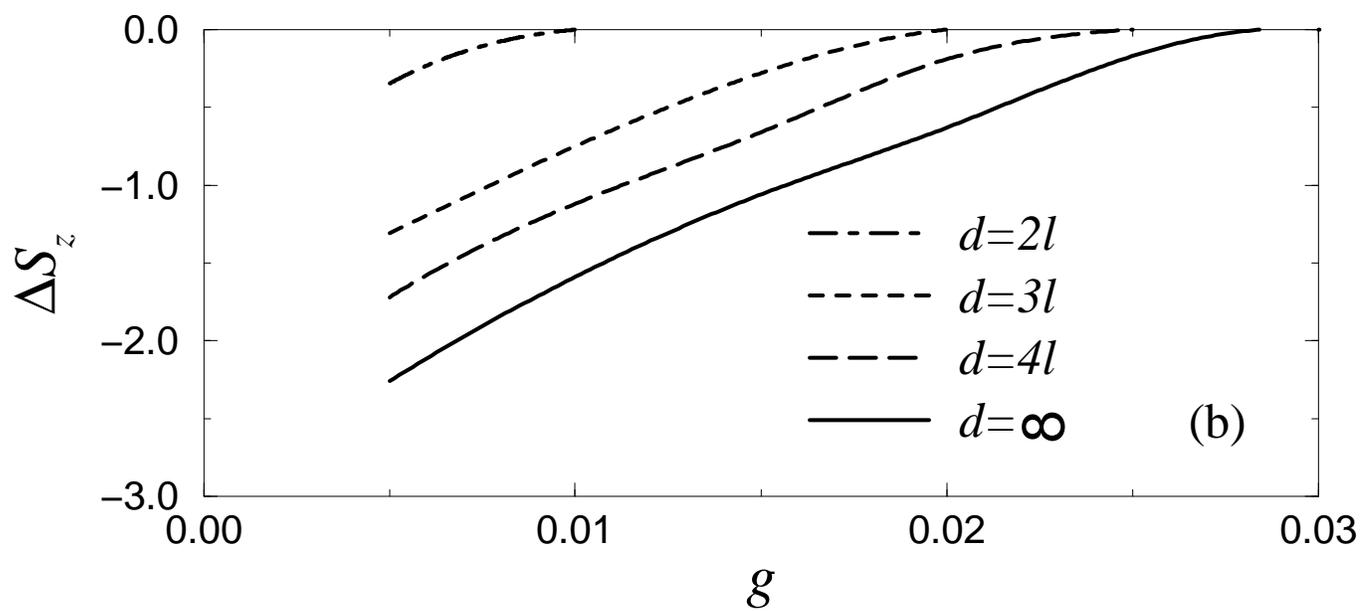
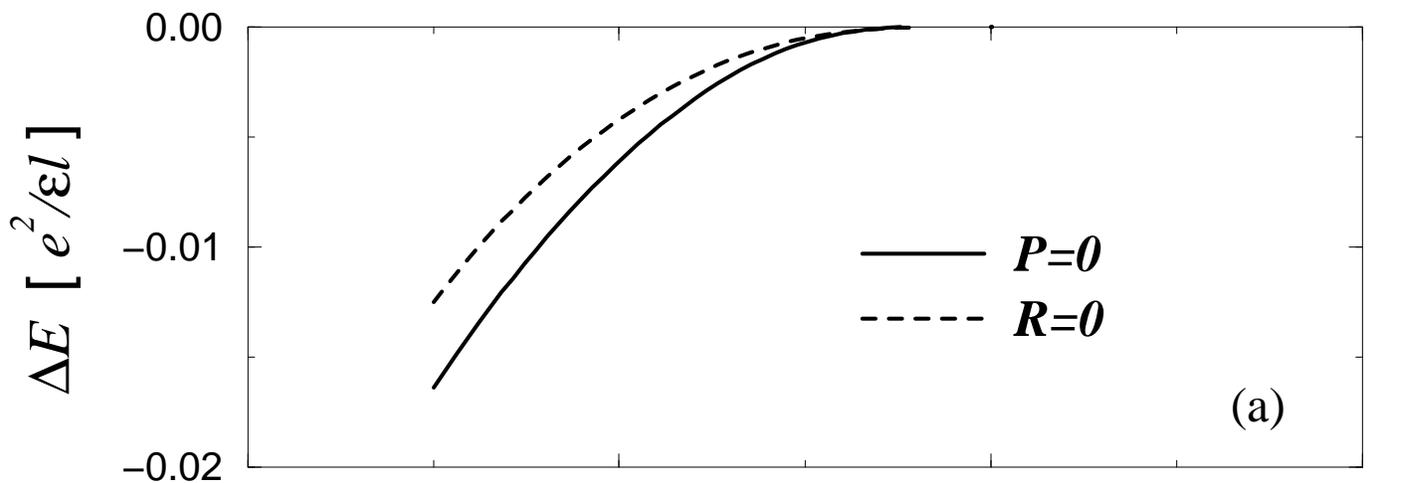
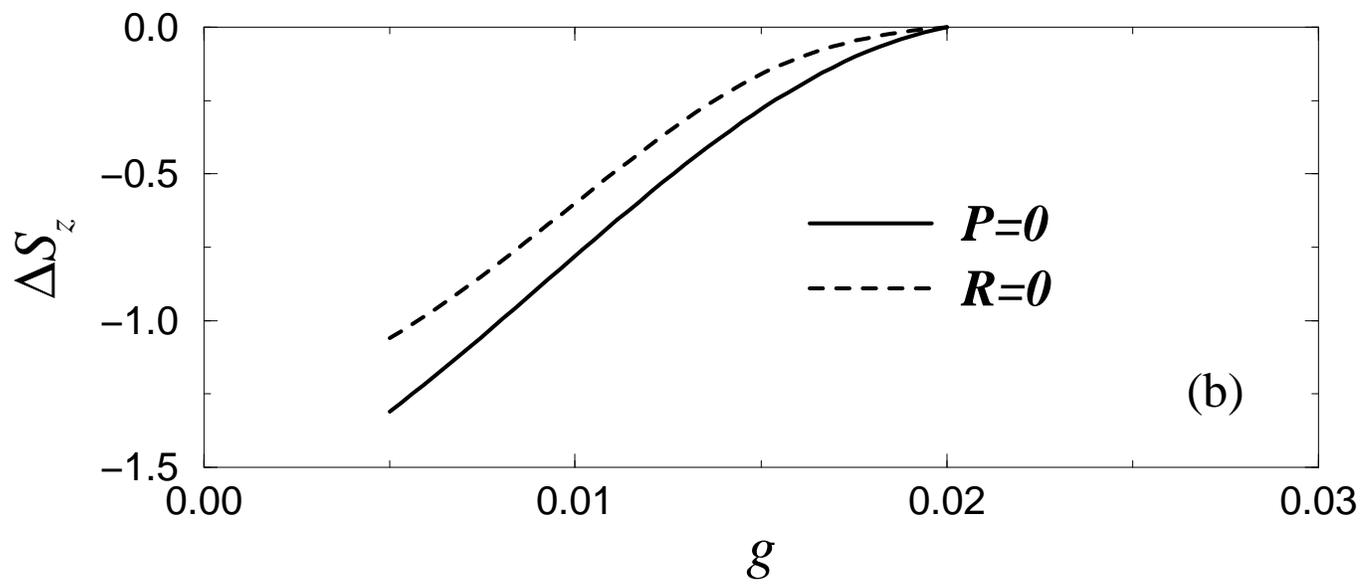
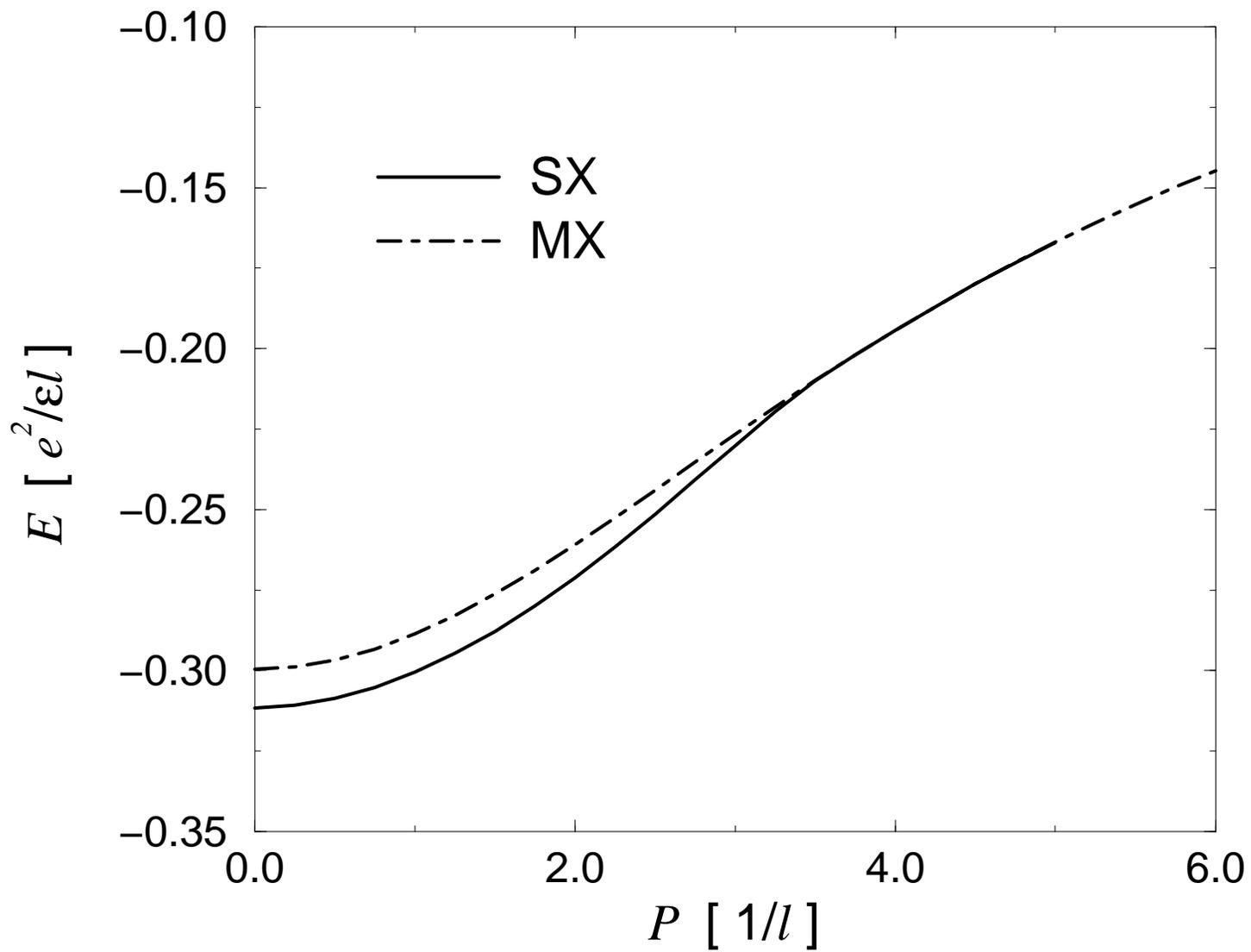
\end{document}